\documentclass[useAMS,usenatbib]{mn2e}
\usepackage{graphicx}
\usepackage{subfig}
\usepackage{longtable}
\usepackage{amsmath}
\usepackage{titlesec}
\usepackage{gensymb}
\usepackage{times}
\def \aj {AJ}
\def \mnras {MNRAS}
\def \pasp {PASP}

\def \apj {ApJ}

\def \apjl {ApJL}
\def \aap {A\&A}

\def \nat {Nature}
\def \araa {ARA\&A}
\def \iaucirc {IAUC}
\def \physrep {Phys. Rep.}

\newcommand{\kms} {$\mathrm{km \; s^{-1}}$}
\newcommand{\msol} {M$_{\odot}$}

\newcommand{\rsol} {R$_{\odot}$}

\def\lesssim{\mathrel{\hbox{\rlap{\hbox{\lower4pt\hbox{$\sim$}}}\hbox{$<$}}}}
\def\gtrsim{\mathrel{\hbox{\rlap{\hbox{\lower4pt\hbox{$\sim$}}}\hbox{$>$}}}}
\newcommand{\halpha} {$\mathrm{H\alpha}$}
\newcommand{\ang} {$\mathrm{\AA}$}

\newcommand{\hena}{He\,{\sc i}/Na\,{\sc i}\,{\sc d}}
\newcommand{\canir}{Ca\,{\sc ii}\,{\sc ir3}}
\newcommand{\heseven}{He\,{\sc i} $\lambda7065$}
\newcommand{\hefive}{He\,{\sc i} $\lambda5876$}
\newcommand{\hesix}{He\,{\sc i} $\lambda6678$}
\newcommand{\naid}{Na\,{\sc i}\,{\sc d}}
\begin{document}
%%%%%%%%%%%%%%%%%%%%%%%%%%%%%%%
%%%%%%%%%%     Title Page    %%%%%%%%%%%%%
%%%%%%%%%%%%%%%%%%%%%%%%%%%%%%%
\title[Spectropolarimetry of the Type Ib Supernova iPTF 13bvn]{Spectropolarimetry of the Type Ib Supernova iPTF 13bvn: Revealing the complex explosion geometry of a stripped-envelope core-collapse  supernova. \thanks{Based on observations made with ESO Telescopes at the Paranal Observatory, under programme 091.D-0516.}}
\author[Reilly et al.]{Emma Reilly$^{1}$\thanks{Email: ereilly528@qub.ac.uk}, Justyn R. Maund$^{2,3}$, Dietrich Baade$^{4}$, J. Craig Wheeler$^{5}$, \newauthor{Jeffrey M. Silverman$^{5}$, Alejandro Clocchiatti$^{6}$, Ferdinando Patat$^{4}$, Peter H{\"o}flich$^{7}$}, \newauthor{Jason Spyromilio$^{4}$, Lifan Wang$^{8}$ and Paula Zelaya$^{6}$}\\
$^{1}\,$Astrophysics Research Centre, School of Mathematics and Physics, Queens University Belfast, Belfast BT7 1NN, UK\\
$^{2}\,$Department of Physics and Astronomy, The University of Sheffield, Hicks Building, Hounsfield Road, Sheffield, S3 7RH, UK\\
$^{3}\,$Royal Society Research Fellow\\
$^{4}\,$ESO - European Organisation for Astronomical Research in the Southern Hemisphere, Karl-Schwarzschild-Str. 2, 85748 Garching b. M{\"u}nchen, Germany\\
$^{5}\,$Department of Astronomy, University of Texas at Austin, Austin, TX 78712-1205, USA\\
$^{6}\,$Pontificia Universidad Cat{\'o}lica de Chile, Avda. Libertador Bernardo O' Higgins 340, Santiago, Chile\\
$^{7}\,$Department of Physics, Florida State University, Tallahassee, Florida 32306-4350, U.S.A.\\
$^{8}\,$Department of Physics, Texas A\&M University, College Station, Texas 77843-4242, U.S.A.\\
}

\maketitle

\begin{abstract}
We present six epochs of spectropolarimetric observations and one epoch of spectroscopy of the Type Ib SN iPTF 13bvn. The epochs of these observations correspond to $-$10 to $+$61 days with respect to the {\it r}-band light curve maximum. The continuum is intrinsically polarised to the 0.2-0.4\% level throughout the observations, implying asphericities of $\sim$10\% in the shape of the photosphere. We observe significant line polarisation associated with the spectral features of \canir, \hena, He\,{\sc i}\,$\lambda\lambda 6678, 7065$, Fe\,{\sc ii}\,$\lambda 4924$ and O\,{\sc i}\,$\lambda 7774$.  We propose that an absorption feature at $\sim 6200\mathrm{\AA}$, usually identified as Si\,{\sc ii}\,$\lambda 6355$, is most likely to be high velocity $\mathrm{H\alpha}$ at $-16,400$\kms.  Two distinctly polarised components, separated in velocity, are detected for both \hena\,\,and \canir\,, indicating the presence of two discrete line forming regions in the ejecta in both radial velocity space and in the plane of the sky.  We use the polarisation of He\,{\sc i}\,$\lambda 5876$ as a tracer of sources of non-thermal excitation in the ejecta;  finding that the bulk of the radioactive nickel was constrained to lie interior to $\sim 50-65\%$ of the ejecta radius. The observed polarisation is also discussed in the context of the possible progenitor system of iPTF 13bvn, with our observations favouring the explosion of a star with an extended, distorted envelope rather than a compact Wolf-Rayet star.

\end{abstract}
\begin{keywords}
supernovae: general - supernovae: individual: iPTF 13bvn - techniques: polarimetric
\end{keywords}

%%%%%%%%%%%%%%%%%%%%%%%%%%%%%%%
%%%%%%%%%%     Introduction  %%%%%%%%%%%%%
%%%%%%%%%%%%%%%%%%%%%%%%%%%%%%%
\section{Introduction}
\label{sec:intro}
The most intriguing and mysterious aspect of the cataclysmic death of massive stars as supernovae (SNe) is the explosion mechanism itself. The driving force of the explosion is the subject of numerous theoretical models (see recent reviews in \citealt{Bur2013} and \citealt{Jan2012}), where the distinguishing feature between the models is the resulting geometry of the explosion.  The vast majority of extragalactic SNe, however, are too distant for their geometries to be directly spatially resolved with currently available, or even planned, telescope facilities. Spectropolarimetry is a powerful probe of the 3D geometries of unresolved SNe and can therefore aid in pinning down the nature of the explosion. 

Previous polarimetric studies have shown that departures from spherical symmetry are present in most if not all SNe \citep{Wan2008}. Thermonuclear Type Ia SN explosions exhibit strong polarisation associated with absorption lines before peak luminosity, which subsequently decrease in strength as the SNe evolve. This implies that the inner ejecta are more spherical than the outer layers. Conversely, core collapse SNe (CCSNe) such as SNe 1993J, 2002ap, 2005bf and 2008D were observed to be highly polarised across particular lines and show asymmetries in the shape of the photosphere on the order of $\sim$10\% \citep{Tra1993, Tran1997, Kaw2002, Wan2003ap, Mau2007b, Mau2009}. In CCSNe, the degree of asymmetry appears to increase as the photosphere recedes into the deeper layers of the ejecta. The Type IIP SN 2004dj showed a dramatic increase in the continuum polarisation as the photosphere receded through the hydrogen layer into the core, unambiguously signalling that the explosion mechanism itself was inherently asymmetric and previously shielded from view by the hydrogen envelope at earlier times \citep{Leo2006}. Polarimetry of SNe arising from progenitors that have been stripped of their outer hydrogen envelope probes the geometry of the explosion at early times.

iPTF 13bvn was discovered by the Intermediate Palomar Transient Factory (iPTF) \citep{Law2009} on June 16.238 2013 in galaxy NGC 5806 (22.5 Mpc) and was classified as a Type Ib SN. According to \citet{Cao2013}, the SN was discovered very early at only 0.57 days after their estimated explosion date of June 15.67 2013. The SN reached a peak maximum brightness of $M_{r}=-16.6$ on July 3 2013 (18 days later). 

iPTF 13bvn is the first Type Ib SN for which a possible progenitor candidate has been detected; however the nature of the progenitor candidate is unclear. \citet{Cao2013} identified a blue object in HST/ACS pre-explosion observations, consistent with a single Wolf-Rayet star. This conclusion was supported by \citet{Gro2013} who constrained the initial mass to 31-35 \msol ($\sim$11 \msol\,\,pre-SN), and predicted a large ejecta mass ($\sim$8 \msol). \citet{Fre2014} and \citet{Sri2014} both constrained the ejected mass to $\sim$2 \msol, which is inconsistent with the explosion of such a massive Wolf-Rayet star. Furthermore, hydrodynamical models of \citet{Ber2014} suggested that the progenitor had a mass of 3.5 \msol\,\,prior to the explosion and proposed an interacting binary as the progenitor channel. Reanalysis of the pre-explosion photometry and comparison with binary evolution models by \citet{Eld2015} also suggested that the progenitor was likely to be a low mass helium giant star (initial mass ~10-20 \msol) in an interacting binary system. 

Spectroscopically, iPTF 13bvn showed similarities to Type Ib SNe 2009jf, 2008D and 2007Y \citep{Sri2014, Fre2014}. The light curve, however, showed a much faster decline than most Type Ib/c SNe and a lower peak brightness luminosity \citep{Sri2014}. Reproduction of the light curve with hydrodynamical models required 0.05-0.1 \msol\,of nickel (consistent with an estimate by \citealt{Sri2014}) to be highly mixed out to the outermost layers of the ejecta to reproduce the rise time \citep[][but note that there may be issues with the treatment of opacity in standard models of the peaks of stripped-envelope SNe, \citealt{Whe2015}]{Fre2014, Ber2014}. 

Here we present multi-epoch spectropolarimetry of iPTF 13bvn acquired with the European Southern Observatory (ESO) Very Large Telescope (VLT), covering from $-$10 to $+$36 days with respect to the {\it r}-band peak luminosity plus an additional epoch of spectroscopy at $+$61 days. The sections are organised as follows: the observations and data reduction are presented in Section 2; the results of the observations are presented and analysed in Sections 3 and 4, respectively. In Section 5 the results and analysis are discussed and in Section 6 we present our conclusions.

%%%%%%%%%%%%%%%%%%%%%%%%%%%%%%%
%%%%%%%%%%     Obs & Red  %%%%%%%%%%%%%
%%%%%%%%%%%%%%%%%%%%%%%%%%%%%%%
\section{Observations and Data Reduction}
\label{sec:obs}
\begin{table}
%%%%%%%%%%%% OBS TABLE %%%%%%%%%%%%%%%%
  \caption{Table of VLT FORS2 observations of iPTF 13bvn}
     \begin{tabular}[alignment]{c c c c c c}
  \hline
  \hline
  Object & Date & Exposure & Mean  & Epoch\dag & S/N$\star$\\
   & (UT) & (s) & Airmass & (days) & \\
  \hline
 
  iPTF 13bvn & 2013 06 23  & $8\times870$ & 1.16 & $-$10 & 410 \\
  LTT 6248\ddag  & 2013 06 23 & 40 & 1.00 &  & \\ 
    \\
   iPTF 13bvn & 2013 07 04 & $8\times855$ & 1.13 & 0 & 740\\
   LTT 6248\ddag  & 2013 07 04 & 80 & 1.02 &  & \\  
   \\
   iPTF 13bvn & 2013 07 10  & $8\times855$ & 1.43 & $+$7 & 640\\
   LTT 6248\ddag  & 2013 07 10  & 45 & 1.40 &  & \\  
   \\
 iPTF 13bvn & 2013 07 12 & $8\times855$ & 1.37 & $+$9 & 660\\
 LTT 6248\ddag  & 2013 07 12 & 45 & 1.01 &  & \\  
   \\
  iPTF 13bvn & 2013 07 28 & $8\times855$ & 1.32 & $+$25 & 350\\
 LTT 6248\ddag  & 2013 07 28 & 90 & 1.1 &  & \\  
   \\
iPTF 13bvn & 2013 08 08 & $8\times855$ & 1.68 & $+$36 &  270\\
LTT 6248\ddag  & 2013 08 08 & 90 & 1.05 &  & \\  
 \\
iPTF 13bvn$\ast$ & 2013 09 02 & $3\times1180$ & 1.79 & $+$61 &  \\
LTT 7379$\ast$\ddag  & 2013 09 02  & 3 & 1.08 &  & \\  

\hline
\hline
  \end{tabular}
  \vspace{2mm}
  
  \begin{flushleft}
  \dag Relative to the {\it r}-band maximum on the 3$^{\mathrm{rd}}$ July 2013, according to \citet{Cao2013}\\
  \ddag Flux standard \\
  $\ast$ Spectroscopic observation\\
  $\star$ Average signal-to-noise ratio measured in the continuum regions (as identified in Figure \ref{fig:wlpol}).
  \end{flushleft}
  \label{obstable}
\end{table}

\begin{figure*}
%%%%%%%Fig 1%%%%%%%%
\includegraphics[scale=0.4]{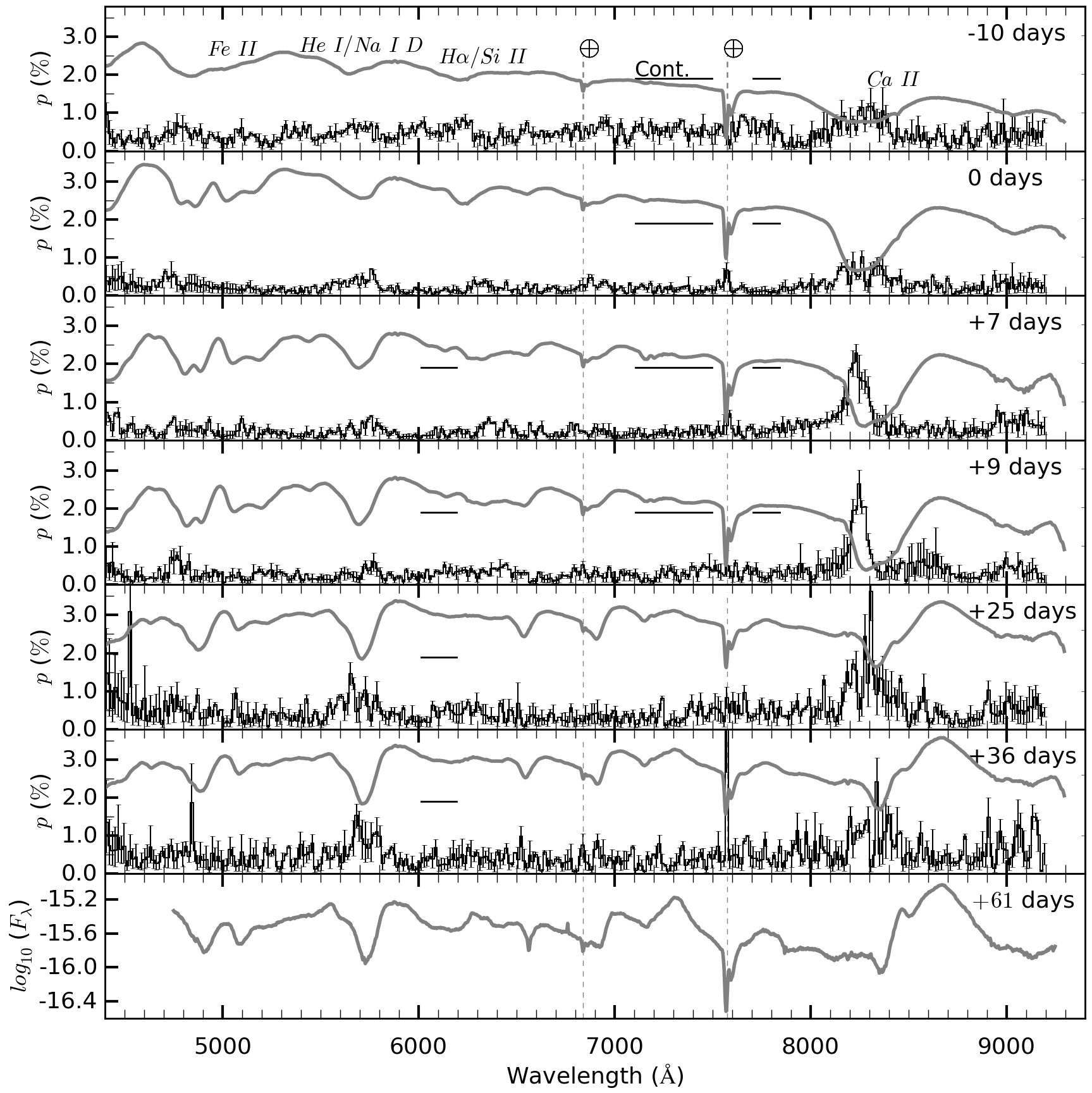}
\caption[]{The polarisation spectra of iPTF 13bvn from $-$10 days to $+$36 days relative to the {\it r}-band maximum. At each epoch, the flux spectra $\left(log_{10}\left(F_{\lambda}\right)\right)$ are shown by the thick grey lines and are corrected for the recessional velocity of the host galaxy ($1359$\,\kms, as quoted by NED). The flux spectrum from the spectroscopic observations at $+$61 days is also shown in the bottom panel. The horizontal bars indicate the wavelength regions used to measure the continuum polarisation.}
\label{fig:wlpol}
\end{figure*}

Observations of iPTF 13bvn were acquired using the  European Southern Observatory (ESO) Very Large Telescope (VLT) Antu (Unit 1) telescope and the Focal Reducer and Low Dispersion Spectrograph (FORS2) instrument in the spectropolarimetric PMOS and long-slit LSS modes \citep{FORS}. Six epochs of spectropolarimetric data and one epoch of spectroscopic data were obtained.  A log of the observations can be found in Table \ref{obstable}.  For the polarimetric observations, the half-wavelength retarder plate was positioned at four angles ($0, 22.5, 45, 67.5^{\circ}$) through two iterations, resulting in eight exposures at each epoch. Observations were acquired with the 300V grism, which provides a spectral resolution of 12 \ang\,\,as measured from arc lamp calibration frames.  The GG435 order separation filter was used to prevent second order contamination at longer wavelengths, resulting in a final wavelength range of 4450-8650 \ang.  The data were reduced in the standard manner using IRAF\footnote{IRAF is distributed by the National Optical Astronomy Observatories, which are operated by the Association of Universities for Research in Astronomy, Inc., under cooperative agreement with the National Science Foundation -http://iraf.noao.edu/.}, following the prescription of \citet{Mau2007b}, and the normalised Stokes parameters ($q$ and $u$) were calculated following \citet{Pat2006}.  The data were corrected for the wavelength dependent chromatic zero-angle offset of the retarder plate, and the polarisation spectra were further corrected for bias following \citet{Qui2012}. In order to increase the signal-to-noise level, the data were rebinned to 15 \ang, prior to the calculation of the Stokes parameters.  Flux spectra of the SN at each epoch of polarimetry were calibrated using observations of a flux standard star with the polarimetry optics in place and the retarder plate at 0$^{\circ}$.
The flux spectrum at $+$61 days was calibrated in the standard manner using observations of the flux standard LTT 7379. 
%%%%%%%%%%%%%%%%%%%%%%%%%%%%%%%
%%%%%%%%%%     Results  %%%%%%%%%%%%%
%%%%%%%%%%%%%%%%%%%%%%%%%%%%%%%
\section{Spectral Evolution}
\label{sec:specev}

The flux spectra and degree of polarisation, $p$, of iPTF 13bvn for the six epochs of spectropolarimetric observations and one epoch of late time spectroscopy are presented in Figure \ref{fig:wlpol}. The phases indicated are with respect to the {\it r}-band maximum on 3 July 2013 \citep{Cao2013}.  

At all epochs, the flux spectra exhibit broad P Cygni profiles of Ca\,{\sc ii}, He\,{\sc i} and Fe\,{\sc ii}. The flux spectra are dominated by a strong absorption due to the Ca\,{\sc ii} near infrared triplet (hereafter Ca\,{\sc ii}\,{\sc ir3}). At $-$10 days, the absorption exhibits a flat bottom, with a minimum at $-$13,300 \kms, which steadily decreases to $-$7450 \kms\,at $+$36 days. At $+$7 days the absorption becomes asymmetric, appearing to be double dipped with apparent minima at $-$10,200 \kms\,and $-$7000 \kms\,, with a blue edge extending to $-$24,000 \kms. By $+$25 days, the two minima occur at $-$4560 \kms\,and $-$8400 \kms.   From $+25$ days another absorption features appears at $\sim$8125 \ang\,, corresponding to $-$16,200 \kms\,with respect to the weighted average rest wavelength of Ca\,{\sc ii}\,{\sc ir3}. Alternatively, this feature may be associated with a possible emission bump observed at 8169 \ang. By $+$61 days, Ca\,{\sc ii} {\sc ir3} is observed to be predominantly in emission, with some evidence that the constituent lines are partially resolved.  The final velocity at the absorption minimum corresponds to either $-$7700 \kms\,(assuming the weighted average rest wavelength for the triplet) or $-$5200 \kms\,(assuming the lowest rest wavelength of the individual Ca\,{\sc ii} {\sc ir3} lines 8498 \ang).

\begin{figure}
%%%%%%%Fig 3%%%%%%%%
\includegraphics[scale=0.285]{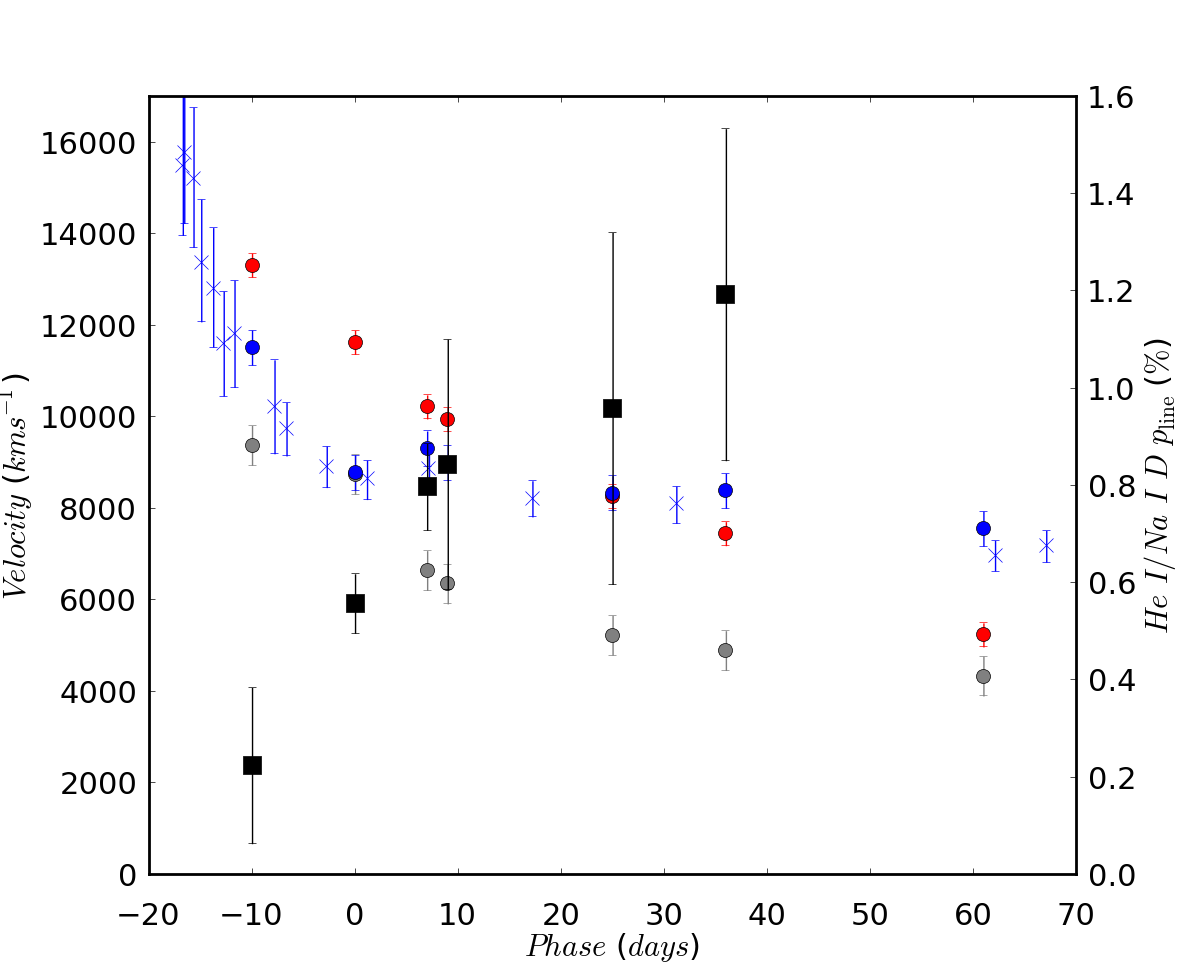}
\caption[]{The evolution of the velocity at absorption minimum for \canir\, (red), the He\,{\sc i}/Na\,{\sc i}\,{\sc d} blend (blue) and Fe\,{\sc ii} $\lambda5169$ (a proxy for the photospheric velocity; grey). These data are marked by circles while measurements for He\,{\sc i}/Na\,{\sc i}\,{\sc d} and Fe\,{\sc ii} $\lambda5169$ from \citet{Fre2014} are also shown here by crosses. The evolution of peak line polarisation associated with the He\,{\sc i}/Na\,{\sc i}\,{\sc d} blend is marked by black squares.}
\label{fig:minabsvel}
\end{figure}

Following the Ca\,{\sc ii}\,{\sc ir3}, the next strongest feature in the spectra is the blend of He\,{\sc i} $\lambda5876$ and Na\,{\sc i}\,{\sc d} at $\sim$5500-5800 \ang. Over the evolution of the SN, the strength of this absorption feature increases while the velocity at the absorption minimum exhibits a slower decline compared to that observed for Ca\,{\sc ii}. The velocity (assuming a rest wavelength of 5876 \ang) decreases from $-$11,500 \kms\,at $-$10 days to $-$8800 \kms\,\,at $0$ days, where it appears to plateau before settling at $-$7900 \kms\,at $+$61 days. The decay in the velocity at absorption minimum for \hena\,, along with \canir\,and Fe\,{\sc ii} $\lambda5169$, is plotted in Figure \ref{fig:minabsvel}. The \hena\,\,line profile appears asymmetric at early times with a blue edge extending up to $\sim-$20,000 \kms, and, at later epochs, there is a possible higher velocity component at $\sim-$14,000 \kms. He\,{\sc i} $\lambda\lambda 6678, 7065$ are also present in the spectra, however obtaining the velocity at absorption minimum for these lines is complicated by blending with a telluric feature for He\,{\sc i} $\lambda7065$ and with the emission component of the Si\,{\sc ii} $\lambda6355$ P Cygni profile for He\,{\sc i} $\lambda 6678$. At $+$25 and $+$36 days, the depth of the He\,{\sc i} $\lambda 7065$ absorption component is greater than that of the telluric lines, such that the velocity at minimum absorption could be measured to be $-$6700 and $-$6500  \kms\,\,at $+$25 and $+$36 days, respectively.

The absorption feature at $\sim 6200\mathrm{\AA}$, observed at $-10$ days, is frequently associated with Si\,{\sc ii} $\lambda6355$.  The strength of the line and the velocity of the absorption minimum decreases with time, eventually disappearing at $+$25 days. The corresponding velocity at absorption minimum of $-$7400 \kms, however, is lower than the photospheric velocity at the same epoch (see below and Fig. \ref{fig:minabsvel}).  Alternatively, \citet{Par2015} suggest that features such as this in other Type I CCSNe may be high velocity $\mathrm{H\alpha}$.  If this feature arises instead from hydrogen the corresponding velocity at absorption at -10 days was $16,700\,\mathrm{km\,s^{-1}}$.

At $-$10 days, Fe {\sc ii} is observed as an unresolved blend of doublets 38 and 42 covering the wavelength region 4800-5400 \ang, before the individual lines become resolved by 0 days. Following \citet{Fre2014}, the velocity at the absorption minimum of Fe\,{\sc ii} $\lambda5169$ is used to indicate the photospheric velocity, for which we measure a velocity of $-$8750 \kms\,\,at 0 days.  The photospheric velocity steadily decreased to $-$4300 \kms\,by $+$61 days.

O\,{\sc i} $\lambda 7774$ appears as a weak P Cygni profile from $+$25 days onwards, having been absent at earlier epochs.

%%%%%%%%%%%%%%%%%%%%%%%%%%%%%%%
%%%%%%%%%%     Analysis       %%%%%%%%%%%%%
%%%%%%%%%%%%%%%%%%%%%%%%%%%%%%%
\section{Analysis of the Polarimetry}
\subsection{Interstellar Polarisation}
\label{sec:analysisisp}
Scattering of photons due to intervening dust in the interstellar medium introduces an additional component to the observed polarisation. The removal of the interstellar polarisation (ISP) is a vital step in correctly interpreting the intrinsic polarisation of the SN.   The correct determination of the ISP is non-trivial and dependent upon the assumption that certain wavelength regions of the SN spectrum are intrinsically unpolarised. One approach is to assume that resonance scattering photons are intrinsically unpolarised \citep{Tra1993}, such as those found in strong emission lines and areas of line blanketing.  Under the assumption that certain regions of the SN spectrum at certain epochs are intrinsically unpolarised and, therefore, representative of the effect of the ISP,  we derive 3 estimates of the ISP:
\begin{description}
\item[$\mathbf{ISP_{A}}$:]{The observed polarisation for intrinsically unpolarised strong resonant scattering emission lines should tend towards the ISP (as the line strength increases). The strongest emission line in the SN spectra is the Ca\,{\sc ii}\,{\sc ir3} at $+$36 days. The Stokes parameters were averaged over a 60 \ang\,range, redward of the emission peak, avoiding the polarisation associated with the blueward absorption trough.  For our first estimate of the ISP, we measured values of the polarisation across this wavelength range of $q_{ISP_A}$=$-$0.28$\pm$0.24$\%$ and $u_{ISP_A}$=0.12$\pm$0.25$\%$.}
\item[$\mathbf{ISP_{B}}$:]{ At $-$10 days the Fe\,{\sc ii} 38 and 42 doublets are unresolved in the spectrum leading to a region of line blanketing which is assumed to be intrinsically unpolarised. Taking the weighted average of Stokes $q$ and $u$ in the wavelength region 5000-5400 \ang, we derive an estimate for ISP$_{B}$ of $q_{ISP_B}$=$-$0.12$\pm$0.12$\%$ and $u_{ISP_B}$=0.39$\pm$0.14$\%$. }
\item[$\mathbf{ISP_{C}}$:]{A further estimate of the ISP was determined by averaging the polarisation of the Ca\,{\sc ii}\,{\sc ir3} emission line and strong emission from a possible Fe\,{\sc ii} and He\,{\sc i} blend (at $\sim$5010 \ang) at $+$25 days. At this epoch the spectrum appears depolarised in these regions, from which we derive ISP$_{C}$ as $q_{ISP_C}$=$-$0.19$\pm$0.04$\%$ and $u_{ISP_C}$=$-$0.10$\pm$0.08$\%$.}
\end{description}
The inverse error weighted average of the three estimates was taken as the principal estimate of the ISP, with the standard deviation as the corresponding uncertainty, resulting in $q_{ISP}$=$-$0.19$\pm$0.08$\%$ and $u_{ISP}$=0.03$\pm$0.28$\%$. The position of the principal and individual estimates of the ISP are shown on the Stokes $q-u$ plane on Figure \ref{fig:qcontucont}.

The total reddening due to Galactic and host dust is E(B\,$-$\,V)$_{\rm{MW}}$=0.0278 \citep{Sch2011} and E(B\,$-$\,V)$_{\rm{host}}$=0.0437 \citep{Cao2013}. Assuming a Serkowski law and Galactic-type dust, this limits the degree of polarisation arising from the intervening dust to be \mbox{$p_{ISP}\leq$\,9(\,E(B\,$-$\,V)$_{\rm{MW}}$\,$+$\,E(B\,$-$\,V)$_{\rm{host}}$)\,=\,0.64$\%$ }. The value we have determined for $p_{ISP}$=  $0.25\substack{+0.22 \\ -0.17}\%$ is therefore consistent with the maximum expected given the reddening towards iPTF 13bvn.

\subsection{Intrinsic Continuum Polarisation}
\label{sec:contpol}

\begin{figure}
%%%%%%%Fig 3%%%%%%%%
\includegraphics[scale=0.37]{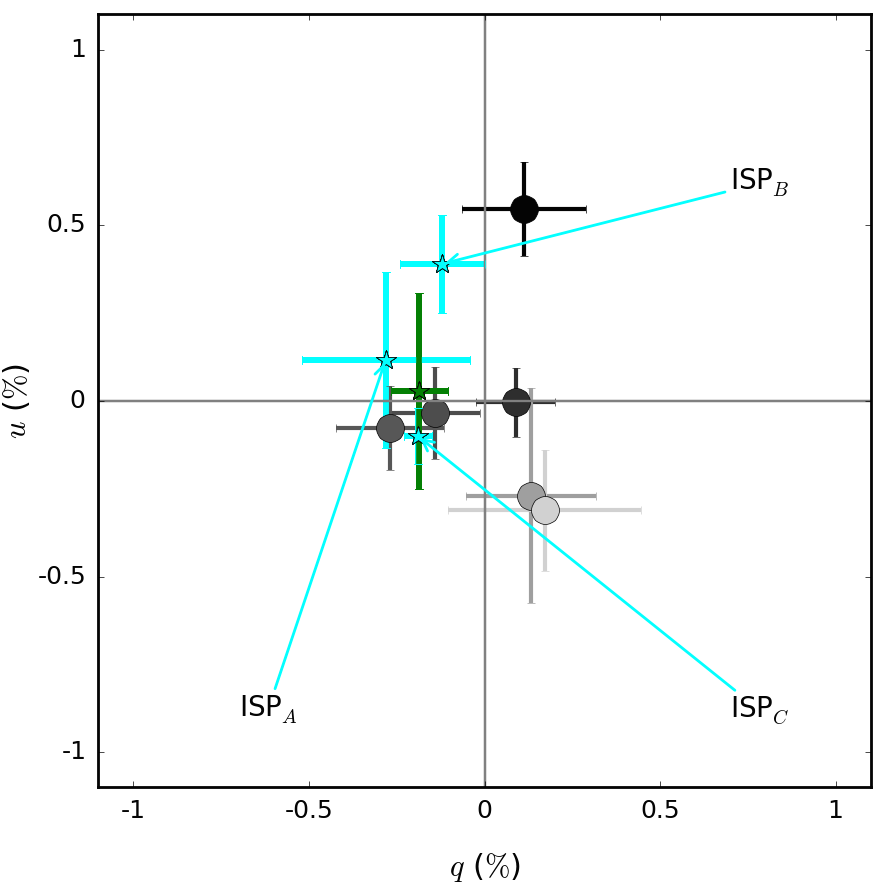}
\caption[]{The temporal evolution of the observed continuum polarisation, coloured according to phase from dark ($-$10 days) to light grey ($+$36 days). The three individual estimates of the ISP (see Section \ref{sec:analysisisp}) are marked by cyan stars and the position of the average ISP estimate is marked by the green star.}
\label{fig:qcontucont}
\end{figure}

The level of intrinsic continuum polarisation reveals the degree of asphericity in the SN photosphere. We identified regions of the spectrum of iPTF 13bvn at each epoch that were deemed to be flat in polarisation and far from any lines that might contaminate the polarisation (indicated by the black horizontal bars in Figure \ref{fig:wlpol}).  As the spectrum of iPTF 13bvn evolved, with lines appearing and disappearing, the wavelength regions we considered to be representative of the continuum also changed. The continuum polarisation was measured by taking the inverse-error weighted-average of the Stokes parameters, after correction for the ISP, over the continuum wavelength ranges. The corresponding uncertainties were calculated as the standard deviation of the polarisation over the continuum wavelengths ranges,  with respect to the weighted mean continuum polarisation. The evolution of the continuum polarisation (before correction for the ISP) is shown in Figure \ref{fig:qcontucont}.

We find the continuum to be intrinsically polarised at the 0.2-0.4$\%$ level throughout the series of observations.  Due to the low level of polarisation, and the relatively large uncertainties ($\sim \pm 0.13\%$), there is little evidence for evolution in the degree of the continuum polarisation to the 3$\sigma$ level.  The continuum polarisation angle undergoes a clockwise rotation from 29$\pm$9$^{\circ}$ to 177$\pm$3$^{\circ}$ between the first two observations. After the second epoch, the polarisation angle remains approximately constant throughout the remaining observations of $+$7, $+$25, $+$36 days at $\sim$153$^{\circ}$; however, at $+$9 days the continuum polarisation angle is 117$\pm$4$^{\circ}$,  despite no change in the degree of polarisation, at the 0.01$\%$ level, from two days before.

The inferred values for the continuum polarisation are, however, sensitive to the choice of the ISP.  For example, correction for $ISP_{A}$ yields similar values for the continuum polarisation as presented above, however the continuum polarisation angle at $+9$ days ($137\pm3^{\circ}$) is no longer discrepant from the angle measured immediately before.  Upon subtraction of $ISP_{B}$, the continuum polarisation exhibits a steady increase from $0.19\pm0.10\%$ before peak brightness to $0.59\pm0.25\%$ at $+36$ days.  The removal of ISP$_{C}$ results in the degree of continuum polarisation reaching a maximum of $\sim$0.7$\%$ at $-$10 days, before decreasing to 0.13$\%$ at $+$7 days and then increasing again to 0.34$\%$ at $+$25 days.  In addition, for ${ISP}_{C}$, the continuum polarisation angle also exhibits very dynamic behaviour changing from $32\pm6^{\circ}$ at -10 days, to $9\pm13^{\circ}$ at 0 days and then $157\pm20^{\circ}$ at $+36$ days.

The maximum intrinsic continuum polarisation of 0.39$\%$, after subtraction of the averaged ISP, implies asphericities of 10$\%$ in the shape of the photosphere \citep{Hof1991}.  For $ISP_{B}$,  the maximum continuum polarisation of 0.7$\%$ implies asphericities of 10-15$\%$.  Although the different ISPs result in different continuum polarisations, the corresponding differences in the inferred shape of the photosphere are relatively small for iPTF 13bvn. The evolution in the alignment of the photosphere on the plane of the sky, however, is shown to be dependent on the chosen ISP. Provided that the ISP is not substantially underestimated in Section \ref{sec:analysisisp} the rotation of the continuum polarisation is intrinsic to the SN.
%%%%%%%%%%%%%%%%%%%%%%%%%%%%%%%%%
%Intrinsic Line Polarisation
%%%%%%%%%%%%%%%%%%%%%%%%%%%%%%%%%
\subsection{Intrinsic Line Polarisation}
\label{sec:linepol}

\begin{figure*}
%%%%%%%Fig 2%%%%%%%%
\includegraphics[scale=0.45]{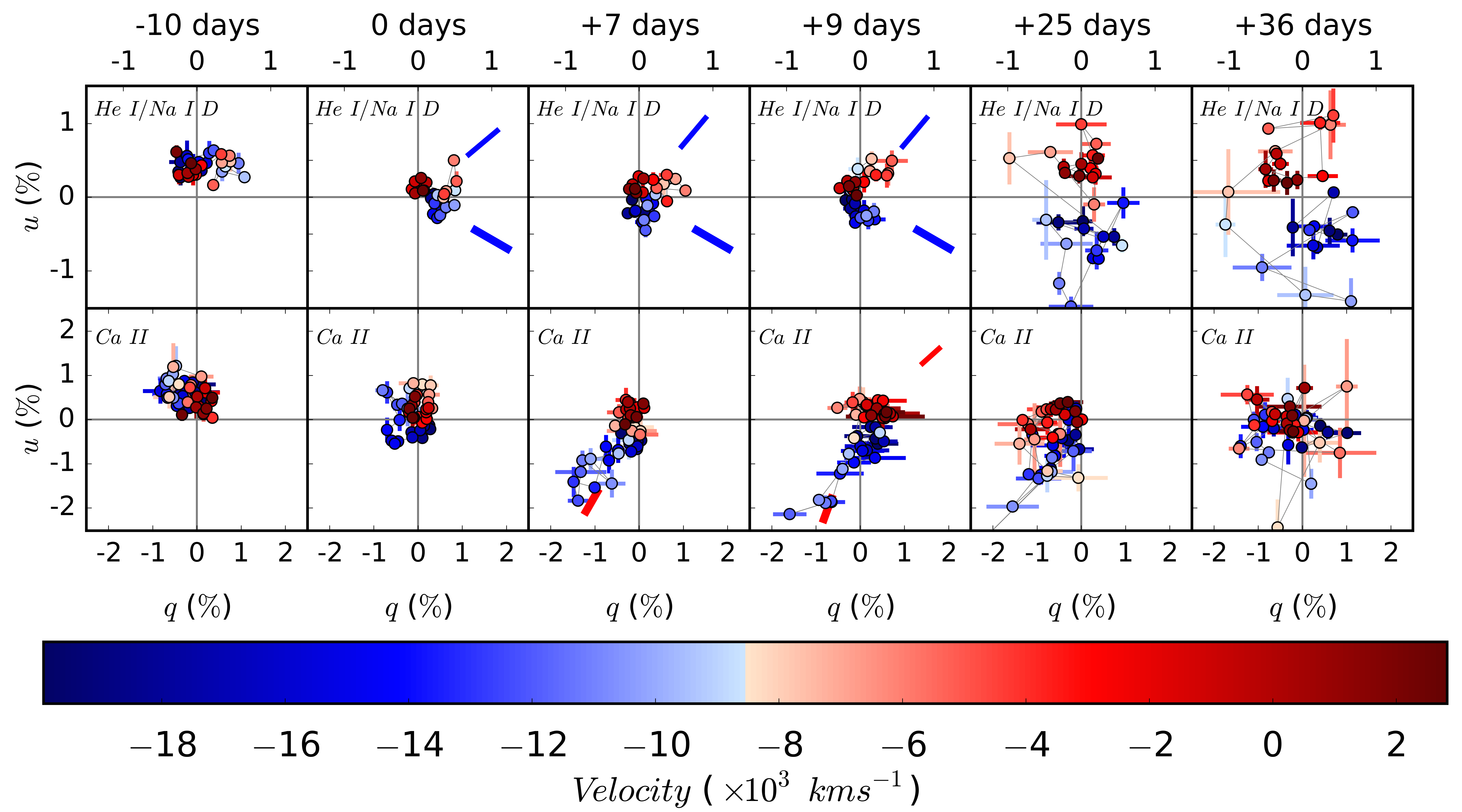}\\
\caption[]{The observed polarisation of the He\,{\sc i}/Na\,{\sc i}\,{\sc d} blend  (top panel) and Ca\,{\sc ii}\,{\sc ir3} (bottom panel) on the Stokes $q$-$u$ plane. The points are coloured coded according their velocity with respect to the rest wavelength.  Radial lines indicate the principal polarisation angles for the LV (narrow lines) and HV (thick lines) components, for epochs at which they appear as separate features.  Note the difference in scale between the top and bottom panels.}
\label{fig:quline}
\end{figure*}

The polarisation spectra, shown in Figure \ref{fig:wlpol}, exhibit peaks associated with the various spectral lines identified in Section \ref{sec:specev}. In the following section we analyse the temporal evolution of the line polarisation intrinsic to iPTF 13bvn.  The sub-sections are organised according to species and ordered from the most strongly polarised species (Ca\,{\sc ii}) to the least (O\,{\sc i} $\lambda 7774$). The intrinsic polarisation due to line absorption features was determined with the vector subtraction of the measured continuum polarisation from the observed data. The observed continuum polarisation is the summation of the intrinsic continuum polarisation of the SN and the ISP and the subtraction of the observed continuum polarisation results, therefore, in an intrinsic line polarisation that is independent of the choice of the ISP.

\subsubsection{\canir}
\label{subsec:Ca}

At all epochs, the polarisation spectra are dominated by strong line polarisation associated with the Ca\,{\sc ii}\,{\sc ir3}. The polarisation signal evolves from a broad feature (FWHM $\sim$9,500 \kms\,) with degree of polarisation $p$=0.7$\pm$0.5\% at $-$10 days  to a narrower, stronger signal of $p\sim$2.5\% at $+$7 days. The polarisation increases again to $p$=3.3$\pm$0.8\% at $+$25 days, with significant signal spread across a larger velocity range than was observed previously. The velocity at which the polarisation across the \canir\, profiles appears to peak also increases from $-$9500 \kms\,\,at $-$10 days to $-$12,300 \kms\,\,at $+$7 days before decreasing again to $-$8400 \kms\,\,at the final epoch of spectropolarimetry. The polarisation angle (at the polarisation maximum) of Ca\,{\sc ii} is similar at $-$10 and 0 days, with $\theta_{\rm line}\sim$70$^{\circ}$. It then rotates to $\sim$120$^{\circ}$ at $+$7 days, where it remains throughout the rest of the observations.   

There is significant structure associated with this line on the Stokes $q$-$u$ plane, shown in Figure \ref{fig:quline}. At 0 days, there is a loop that crosses all quadrants of the $q$-$u$ plane, with 3 separate components at $\sim-$8000, $-$12,000 and $-$15,000 \kms, suggesting a complex line forming structure in the plane of the sky. The shape of the loop changes at $+$7 days, resulting in an extended loop for velocities between $\sim-$7000 \kms\,\,and $\sim-$15,000 \kms, while at lower velocities the signal appears to cluster near the origin of the $q$-$u$ plane and at higher velocities ($\lesssim-$15,000 \kms) the data are clustered at $\sim$135$^{\circ}$. At $+$9 and $+$25 days the feature on the $q$-$u$ plane is similar to that observed $+$7 days, although with a higher degree of polarisation $p\sim$3.3$\%$ seen at $+$25 days. There is evidence of significant polarisation at $+$36 days, however the signal-to-noise ratio is insufficient to properly discern any structure in the $q$-$u$ plane at this epoch.

\begin{figure}
%%%%%%%Fig 4%%%%%%%%
\includegraphics[scale=0.26]{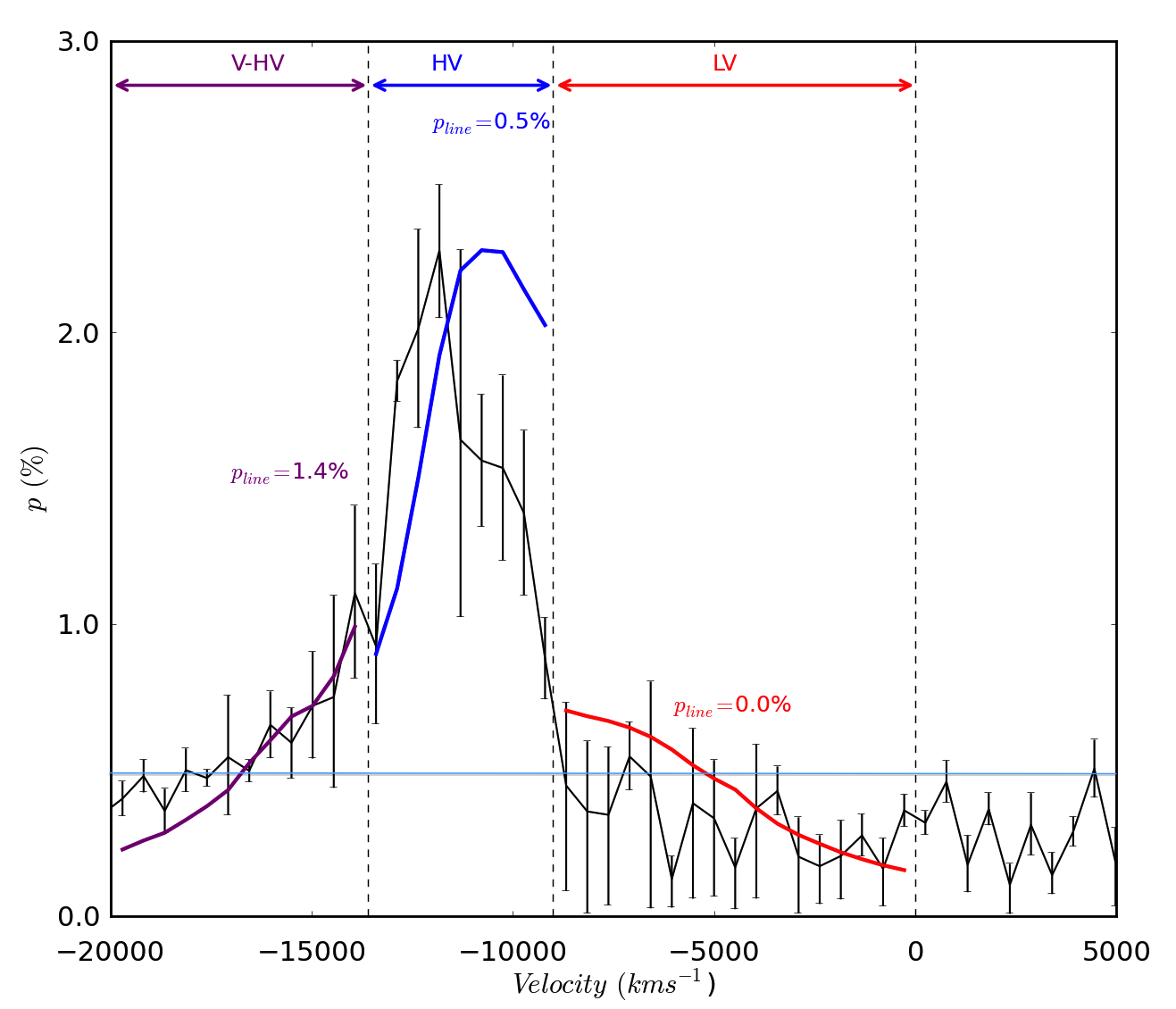}

\caption[]{Fits to the observed degree of polarisation for the Ca\,{\sc ii}\,{\sc ir3} at $+$9 days with $p_{\rm{line}}$ as a variable. No one value of $p_{\rm{line}}$ can provide a complete fit to the observed polarisation across the entire line profile. The best fits to LV, HV and V-HV components of the line are marked in red, blue and purple, respectively.}
\label{fig:capline}
\end{figure}

At $+$7 days, there is evidence of polarised flux in a blue wing to the absorption profile, that extends to velocities of $\sim-$20,000 \kms. To determine if the blue wing is polarised in the same manner as the  strong polarisation seen at $-12,300$ \kms, or due to a separate polarising component, the observed degree of polarisation ($p_{\rm obs}$) was modelled as:
\begin{equation}
p_{\rm obs}=\frac{p_{\rm line}F_{\rm line}+p_{\rm cont}F_{\rm cont}}{F_{\rm obs}}
\label{eqn:pol}
\end{equation}
where $F_{\rm line}$ and $F_{\rm cont}$ are the line and continuum fluxes, respectively, and $p_{\rm cont}$ is the continuum polarisation. These values were measured directly from the observed data.  The intrinsic line polarisation, $p_{\rm line}$, was varied according to Equation \ref{eqn:pol} in order to reproduce $p_{\rm obs}$.  Below $\sim -8000\,$\kms, the observed polarisation is best replicated with  low intrinsic polarisation, $p_{\rm line}$=0.0$\pm$0.2\%, however this is not representative for the data at higher velocities. Between $-$8000 \kms\,\,and $-$14,000 \kms, $p_{\rm line}$=0.6$\pm$0.4\%.  For velocities higher than $\sim -14,000\,$\kms, a higher degree of polarisation is required with $p_{\rm line}$=1.6$\pm$0.2\%. As can be seen in Figure \ref{fig:capline}, the data could not be reproduced with a single $p_{\rm line}$ covering the entire wavelength range of the profile, indicating that a single geometrical configuration is not responsible for the polarisation profile. This suggests that the \canir\,\,line profile actually arises from 3 different structures in the plane of the sky: a weakly polarised, strongly absorbing low velocity (LV) component; a more asymmetric, strongly absorbing, high velocity (HV) component; and a very strongly polarising, but weakly absorbing, very high velocity (V-HV) component. 

The behaviour of the polarisation angle across the line profile displays an intriguing evolution. At $-$10 days, the polarisation remains constant as a function of velocity at $\sim$70$^{\circ}$. By maximum light, however, we observe a steady rotation in the polarisation angle from $\sim$50$^{\circ}$ to 150$^{\circ}$ with increasing velocity. At $+$7 days the polarisation angle across the line profile begins to split into two separate components, with the separation between the two becoming clearly apparent by $+$9 days . At $+$9 days there is a defined break in the polarisation angle at $-$8100 \kms\,\,between the two components (see Figure \ref{fig:thetaline}). The LV component, at $\theta_{\rm line}$=21$^{\circ}$, is separated by $\sim$75$^{\circ}$ (or -55$^{\circ}$) from the HV component (from $-$8100 \kms\,\,to $\sim-$15,000 \kms) at $\theta_{\rm line}$=125$^{\circ}$.  At $+$25 days, the polarisation angle changes steadily from approximately 50$^{\circ}$ at low velocities to 120$^{\circ}$ at $-$10,000 \kms. The data are noisier at $+$36 days, however the trend in the polarisation angle appears to be similar to that observed at the previous epoch.

While separate HV and LV components are observed for Ca\,{\sc ii}\, in the polarised flux spectrum (at $+9$ days), an analysis of the \canir\, line profile in the flux spectrum, following the prescription of \citet{Sil2015}, was not able to recover the two separate components. The asymmetric line profile observed in the flux spectrum, from $+$7 days onwards, was found to result from a superposition of the individual components of the triplet (at $\sim-$8800 \kms) becoming partially resolved.  We note, however, that this analysis assumes that the underlying line profiles are well described by Gaussian profiles.   

The fits also indicated that an additional narrow absorption feature at $+25$ days, at $\sim$8150 \ang\, or $-$14,000 \kms relative to the \canir\, rest wavelength, is probably unrelated to Ca and is probably due to another element.   Although not observed at earlier epochs, the velocity of this feature places it in the ``blue wing" or V-HV component of the \canir\, absorption. This could suggest that the extreme line polarisation measured for the V-HV component at  $+$9 days may not actually be due to Ca.  Conversely, we note that if this separate species were responsible for the polarisation of V-HV component, the degree of polarisation would appear to be anti-correlated with the increasing strength of the line in the flux spectrum at later epochs.  The presence of two components in the \hena\, profiles (discussed in Section \ref{subsec:analysishe}) leads us to conclude that the separate LV and HV components Ca\,{\sc ii} are real. 

\subsubsection{He\,{\sc i} $\lambda \lambda5876,\,6678,\,7065$ and Na\,{\sc i} D}
\label{subsec:analysishe}

Polarisation is also associated with the He\,{\sc i} absorption lines, most prominently seen in the He\,{\sc i}/Na\,{\sc i}\,{\sc d} blend. The intrinsic degree of polarisation of the \hena\, blend grows from $p_{\rm line}$=0.29$\pm$0.06\%  at $-$10 days to 1.2$\pm$0.3\% at the final epoch of polarimetry.  At $-$10 days, the data points of the He\,{\sc i}/Na\,{\sc i}\,{\sc d} blend cluster around the position of the continuum polarisation on the Stokes $q$-$u$ plane (see Figure \ref{fig:quline}). By 0 days, the points begin to separate into two polarised clusters, at $\sim$20$^{\circ}$ and  $\sim$165$^{\circ}$ for  velocities $\gtrsim-$8000 \kms\,and $\lesssim-$8000 \kms, respectively. 

At earlier epochs, the polarisation peak occurs redward of the absorption minimum, at a velocity of  $-$6200 \kms\, between $-$10 and $+$7 days. The velocity at which the polarisation maximum occurs dramatically increases at later times to $-$11,900 \kms\, and $-$10,200 \kms\,at $+$25 and $+$36 days, respectively. The significant increase in the velocity of the maximum degree of polarisation is accompanied by a rotation in the polarisation angle from $\theta_{\rm line}$=20$\pm$8$^{\circ}$ at $+$9 days to $\theta_{\rm line}$=128$\pm$9$^{\circ}$ at $+$25 days. The increase in velocity and the rotation of the polarisation angle at $+$25 days appears to be the result of a HV component becoming more strongly polarised than the LV component (that dominated the polarisation at earlier epochs) as can be seen on Figure \ref{fig:wlpol} and Figure \ref{fig:quline}.   

\begin{figure}
%%%%%%%Fig 5%%%%%%%%
\includegraphics[scale=0.25]{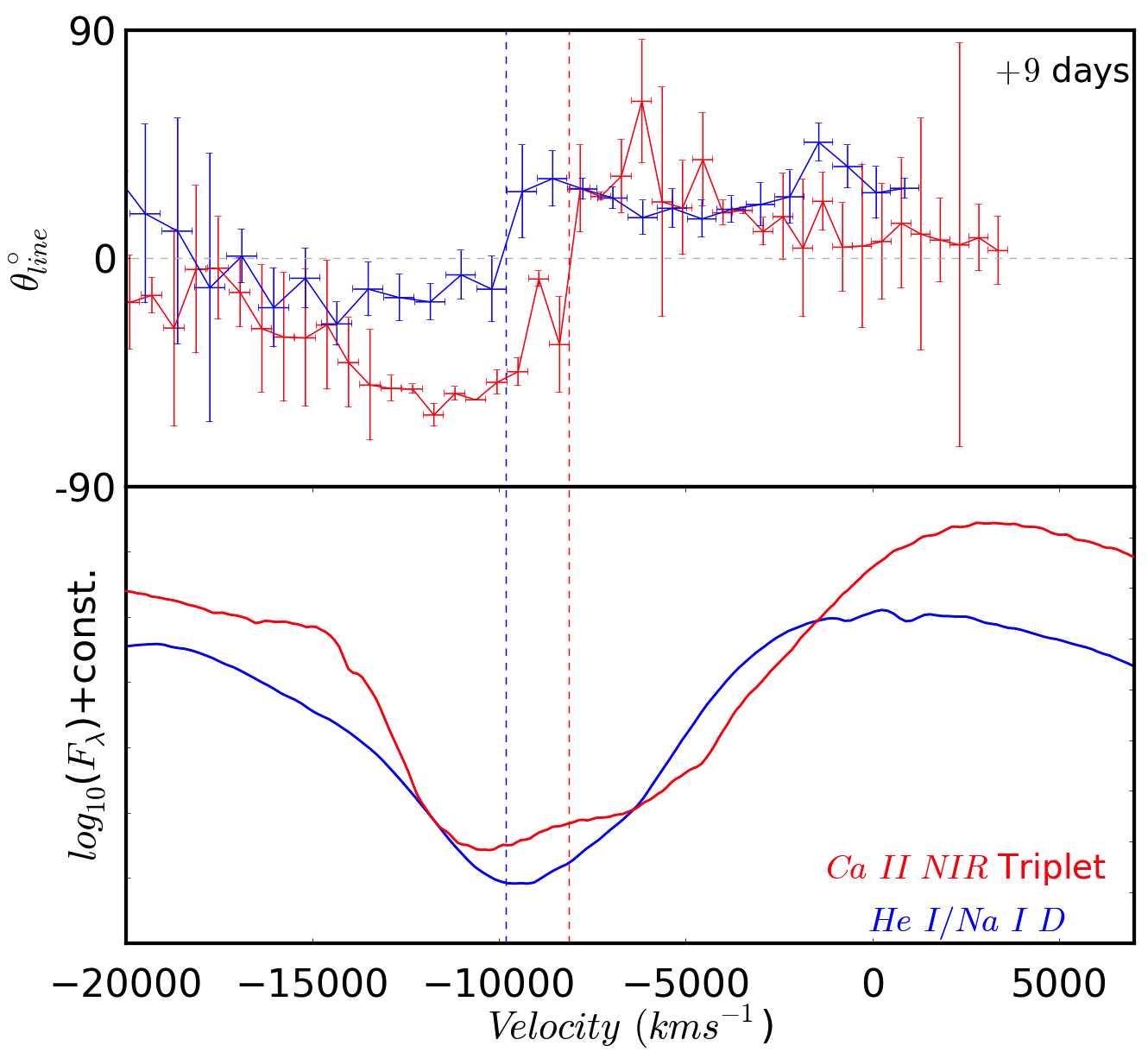}\\
\caption[]{The polarisation angle as a function of velocity for the Ca\,{\sc ii}\,{\sc ir3} (red) and \hena\,\,(blue) profiles at $+$9 days.  The bottom panel shows the total flux spectra across each of these profiles. The dashed lines represent the velocity of the transition between the LV and HV components for \canir\,\,(red, $-$8100 \kms) and \hena\,\,(blue, $-$9800 \kms).}
\label{fig:thetaline}
\end{figure}

As was observed for Ca\,{\sc ii}, the polarisation angle for He\,{\sc i}/Na\,{\sc i}\,{\sc d} blend also exhibits a discontinuity at $+7$ and $+9$ days (see Figure \ref{fig:thetaline}).  This discontinuity occurs at $-$9800 \kms\,, which is higher than the velocity of the discontinuity observed for Ca\,{\sc ii}. For He\,{\sc i}/Na\,{\sc i}\,{\sc d}, the LV component has a polarisation angle of $\theta_{\rm line}$=25$^{\circ}$ and is separated by 40$^{\circ}$ from the HV component at $\theta_{\rm line}$=165$^{\circ}$ (or -15$^{\circ}$).  At $+$25 days, the polarisation angle is contant at $\sim 47^{\circ}$ for velocities less than $-$6500 \kms, before increasing to 141$^{\circ}$ at $-$14,000 \kms. A similar trend is observed at $+$36 days. 

For \hesix\,, significant polarisation is only observed at the final epoch; with $p_{\rm line}$=1.0$\pm$ 0.3\% and  $\theta_{\rm line}$=180$\pm$9$^{\circ}$.  Conversely, polarisation is observed for \heseven\,  at 0, $+$9 and $+$36 days. The degree of polarisation of \heseven\, reaches a maximum at $+$9 days with $p_{\rm line}$=0.73$\pm$0.06\% and $\theta_{\rm line}$=10$\pm$2$^{\circ}$, before undergoing a rotation in the polarisation angle to $138\pm14^{\circ}$ at $+$36 days.   The lack of significant polarisation across the \heseven\,  and \hesix\,lines at all epochs makes it difficult to determine to what degree the Na\,{\sc i} doublet contributes to the polarisation of the \hena\,blend. The shape of the \hefive\, and \heseven\, line profiles show good agreement at 0, $+$7 and $+$9 days, which may indicate a minimal contribution from Na at these epochs. At $-$10, $+$25 and $+$36 days, however, the line profiles are different, with the velocity at the \hefive\, absorption minimum being substantially higher.  The agreement between the \hena\, and \heseven\, polarisation angles at low velocities ($\gtrsim -4000$ \kms) suggests He dominates the polarisation signal in that velocity range.   

We conclude that the HV He\,{\sc i} component, although weakly polarised and potentially contaminated by the polarised Na\,{\sc i} doublet, is a real feature and represents a separate line forming region for He\,{\sc i} at high velocities.  Contamination from polarised \naid\, would result in a steady rotation of the polarisation angle, rather than the discontinuity observed in Figure \ref{fig:thetaline}.   

\subsubsection{Fe\,{\sc ii}}
Peaks in the polarisation are also observed around 4700-4800 \ang, coincident with an Fe\,{\sc ii} absorption trough. As such, we associate it with the nearest redward strong emission of Fe\,{\sc ii} $\lambda4924$.  The strength of the polarisation grows from $p$=0.54$\pm$0.26\% at $-$10 days to $p$=1.4$\pm$1.0\% at $+$36 days. The uncertainty on the last measurement is large and the feature appears to be due to one elevated bin at $-$5000 \kms (which may indicate that the feature at +36 days is an artefact from the reduction). The Fe\,{\sc ii} $\lambda4924$ feature is polarised to the $\sim$0.6\% level for the majority of the observations. The velocity of the maximum polarisation initially increases from $-$9800 \kms\,\,to $-$13,800 \kms\,\,between $-$10 and 0 days, before decreasing back to $-$9800 \kms\,\,at $+$9 days and then increasing again to $-$11,800 \kms\,\,at $+$25 days. The polarisation angle rotates from $\theta_{\rm line}$=83$\pm$14$^{\circ}$ in the first epoch to $\sim$135$^{\circ}$ at 0 days; it then remains approximately constant throughout the remainder of the observations.

Given the significant polarisation associated with Fe\,{\sc ii} $\lambda4924$ at $-$10 days, it is likely that the 5000-5400 \ang\,\,wavelength region, used to estimate $ISP_{B}$, is contaminated with polarisation associated with the absorption troughs of the overlapping Fe\,{\sc ii} lines in the region. Therefore, it is possible that $ISP_{B}$ is a poor estimate of the interstellar polarisation. Upon removal of $ISP_{B}$ from consideration, the inverse error weighted average of the ISP remains unchanged within the quoted uncertainties.

\subsubsection{The $\mathrm{6200 \AA}$ Feature}
Significant polarisation is associated with the absorption at $\sim 6200 \mathrm{\AA}$ in the first two epochs, with a maximum polarisation of $p$=0.59$\pm$0.06\% at $-$10 days. If this feature is due to Si\,{\sc ii}, the velocity of the polarised feature decreases from $-$6400 \kms\, to $-$3500 \kms\, between $-$10 and 0 days.  Alternatively, if the feature is $\mathrm{H\alpha}$ then the velocities at the two epochs are, instead, $-16,400\,\mathrm{km\,s^{-1}}$ to $-13,400\,\mathrm{km\,s^{-1}}$. Between $-10$ and $0$ days, there is also a small rotation in the polarisation angle from $\theta_{\rm line}$=76$\pm$3$^{\circ}$ to $\theta_{\rm line}$=63$\pm$7$^{\circ}$. Significant polarisation is not observed after 0 days, when the strength of the absorption feature decreases until it disappears by $+$25 days. 

\subsubsection{O\,{\sc i}}
O\,{\sc i} $\lambda 7774$ appears in the spectrum as a weak P Cygni profile from $+$25 days onwards. Upon removal of the continuum polarisation O\,{\sc i} remains significantly polarised at $+$36 days. We estimate a polarisation associated with O\,{\sc i} of 0.66$\pm$0.13\% at $-$3800 \kms at $+36$ days, with $\theta_{\rm line}$=154$\pm$6$^{\circ}$; however, this measurement is complicated by the proximity of the line to a telluric feature. 

\subsection{The Polarimetric Evolution in the Plane of the Sky}

\begin{figure*}
     \caption{Polar plots for iPTF 13bvn from $-$10 to $+$36 days relative to the {\it r}-band light curve maximum. The diagrams show $\theta_{\rm line}$ as a function of the velocity (\kms; increasing radially). The polarisation angle across the line profiles for Ca\,{\sc ii}, \hena\,\,and Fe\,{\sc ii} is plotted, with the data at maximum polarisation marked with the heavy line.  For other lines, just the polarisation angle and velocity at the maximum polarisation is plotted. We only show those line features that are significantly detected above the continuum level ($p > p_{cont} + 3\sigma)$.  The length of each arc represents the $\pm$1$\sigma$ uncertainties on the polarisation angle. The photospheric velocity as measured by the velocity at the absorption minimum of Fe\,{\sc ii} $\lambda5169$ is indicated by the black dashed semi-circle, while the continuum polarisation angle is shaded in grey.}
      \includegraphics[]{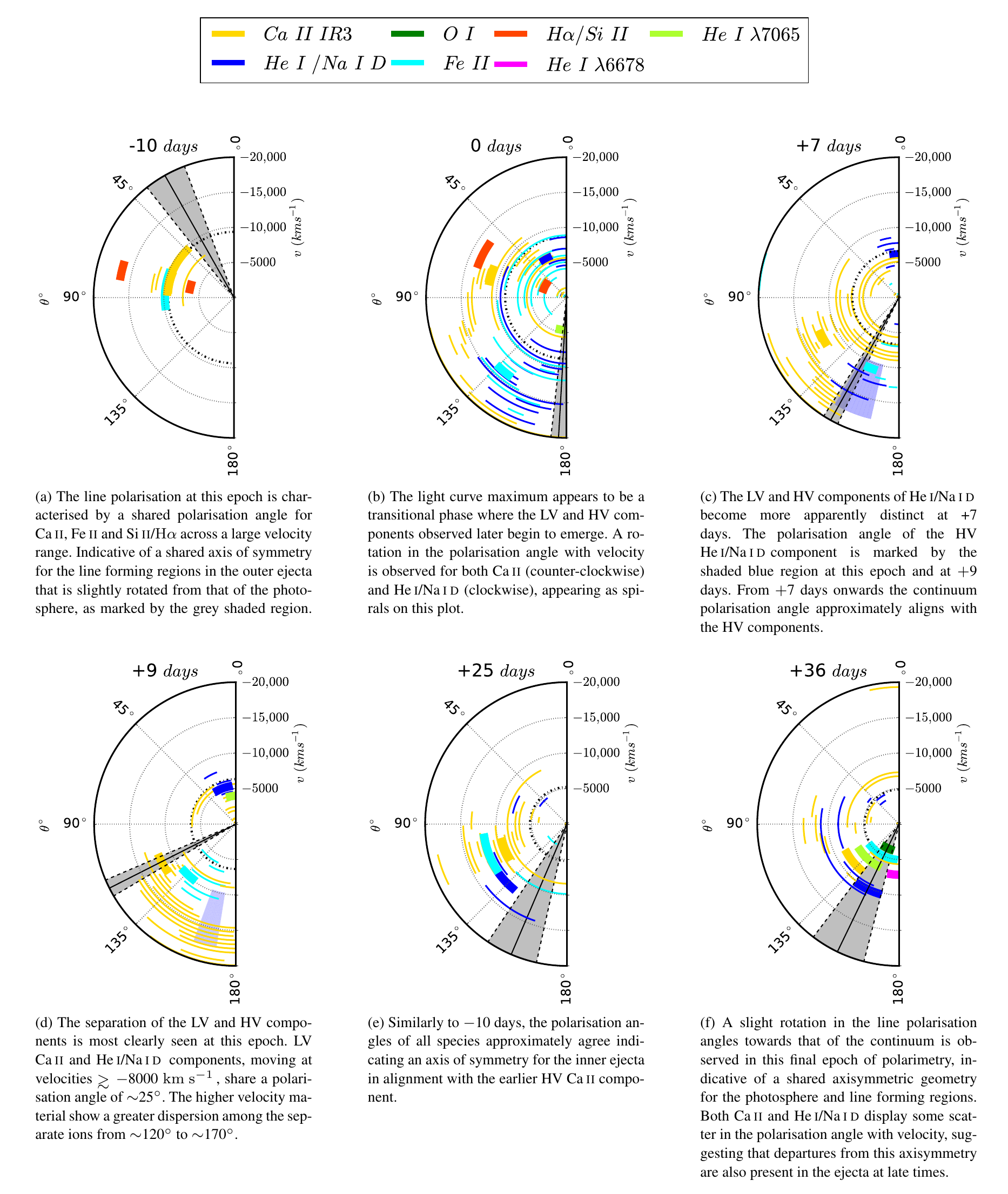}
    \label{fig:polar}

\end{figure*}

The measured polarisation angles provide an indication of the locations of the line forming regions for different species in the plane of the sky.  \citet{Mau2009} suggested a new type of polar plot for spectropolarimetric data, that may be used to illustrate the relative positions of the different line-forming regions in radial velocity space and in the plane of the sky.  These polar plots, or Maund diagrams, provide a projection of the approximate 3D structure of the SN ejecta independently of models.  The spectropolarimetric observations of iPTF 13bvn are presented in this form on Figure \ref{fig:polar}.

The continuum polarisation angle is seen to rotate from $\sim$20$^{\circ}$, before maximum light, by $\sim90^{\circ}$ after maximum (i.e. moving from the top half to the lower half of the polar plot). This is indicative of a substantial change in the orientation of the asymmetry of the photosphere and hence in the underlying excitation structure powering the continuum before and after the light curve maximum. 

At $-$10 days, the most significantly polarised features belong to Ca\,{\sc ii}, Fe\,{\sc ii} and the $6200 \mathrm{\AA}$. The separate species share a polarisation angle on the sky of $\theta_{\rm line}\sim$70$^{\circ}$ which is misaligned with the continuum polarisation angle by $\sim$50 degrees.  

At 0 days a continuous anti-clockwise rotation in the polarisation angle with increasing velocity is observed across the \canir\,\,line profile, creating a spiral-like structure in the polar plot. Conversely, for the He\,{\sc i}/\,Na\,{\sc i}\,{\sc d} blend, the polarisation angle rotates clockwise across the line profile. It wraps around zero degrees (equal to a complete 180$^{\circ}$ rotation of the Stokes parameters) at approximately $-$8000 \kms\,and settles at 140-150$^{\circ}$ for velocities between $-$10,000 and $-$15,000 \kms.  At low velocities ($\gtrsim-$5000 \kms) the polarisation angles for Ca\,{\sc ii} and the He\,{\sc i}/Na\,{\sc i}\,{\sc d} blend, as well as that of the Si\,{\sc ii} $\lambda6355$/$\mathrm{H\alpha}$, are similar at $\theta_{\rm line}\sim$60$^{\circ}$. At higher velocities, the polarisation angles for the two lines diverge and between $-$10,000 and $-$15,000 \kms\,they are offset by $\sim$70 degrees. This suggests that at low velocities the He and Ca line-forming regions occupy a similar location in the ejecta, but not at high velocities.

At $+7$ and $+9$ days, the  two \hena\,\,components are still visible in the polar plot.  The weighted average of the polarisation across the HV component is shaded in blue on the polar plot, where the width of the shaded region represents the standard deviation. The continuum polarisation angle at $+$7 days is aligned with Fe\,{\sc ii} and the HV ($v\lesssim-$9,800 \kms) portion of the \hena\,\,blend. At the same velocities, however, Ca\,{\sc ii} does not share the same orientation, being offset by $\sim$40$^{\circ}$. At $+$7 days there is also some evidence for a further $\sim$30$^{\circ}$ rotation between Ca\,{\sc ii} at $\sim-8000\,-\,15,000$\kms\, and the V-HV Ca\,{\sc ii} component at velocities $\lesssim-$15,000 \kms\, (which tends towards the continuum polarisation angle).  At $+9$ days, the Ca\,{\sc ii} line profile is now also clearly composed of two (possibly three) separate components on the polar plot. The LV components of the He\,{\sc i} $\lambda7065$ and the He\,{\sc i} $\lambda5876$/Na\,{\sc i}\,{\sc d} blend and Ca\,{\sc ii} are all aligned at $\sim 25^{\circ}$.  At this epoch the polarisation angle of Fe\,{\sc ii} has rotated to $\sim$140$^{\circ}$, positioned in between Ca\,{\sc ii} and \hena\,\,on the polar plot. The intrinsic continuum polarisation angle at this epoch also appears to approximately align with that of the HV Ca\,{\sc ii} and Fe\,{\sc ii} components.

Following the decline in the signal-to-noise at later epochs, the data appear more dispersed on the polar plots in the final two epochs of polarimetry ($+$25 and $+$36 days).  At $+$25 days, the LV Ca\,{\sc ii} component appears to undergo an anti-clockwise rotation from $\sim$20$^{\circ}$ at $+$9 days to $\sim$70$^{\circ}$ at $+$25 days, while the HV component appears to be unchanged. At $+$25 days, the most strongly polarised features in the Fe\,{\sc ii}, Ca\,{\sc ii} and He\,{\sc i}/\,Na\,{\sc i}\,{\sc d} line profiles all share the same polarisation angle of $\theta_{\rm line}\sim$120$^{\circ}$ and have similar velocities of $\sim-$10,000 \kms. These features are offset from the continuum orientation by $\sim$40$^{\circ}$.     

At $+$36 days, the polarisation angles of  Fe\,{\sc ii}, Ca\,{\sc ii}, He\,{\sc i} and  O\,{\sc i} $\lambda 7774$ are approximately aligned with that of the continuum at $\sim$160$^{\circ}$.  The alignment of most of the separate species in these later epochs implies that, to some degree, the line-forming regions are shared between the different species in the plane of the sky.

\subsection{Monte Carlo Simulation of the Geometry}
 
\begin{figure}
\centering
%%%%%%%Fig 4%%%%%%%%
\includegraphics[scale=0.425]{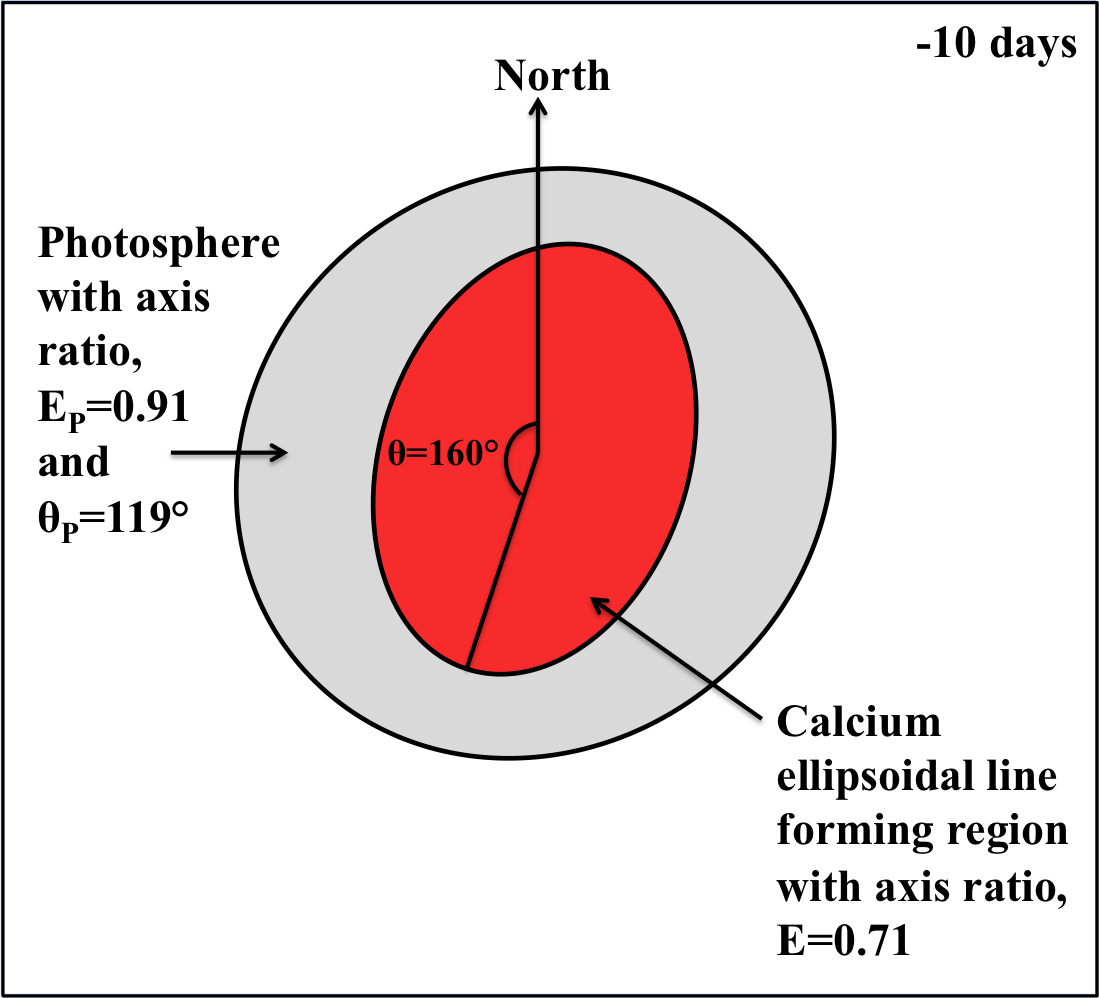}\\
\includegraphics[scale=0.425]{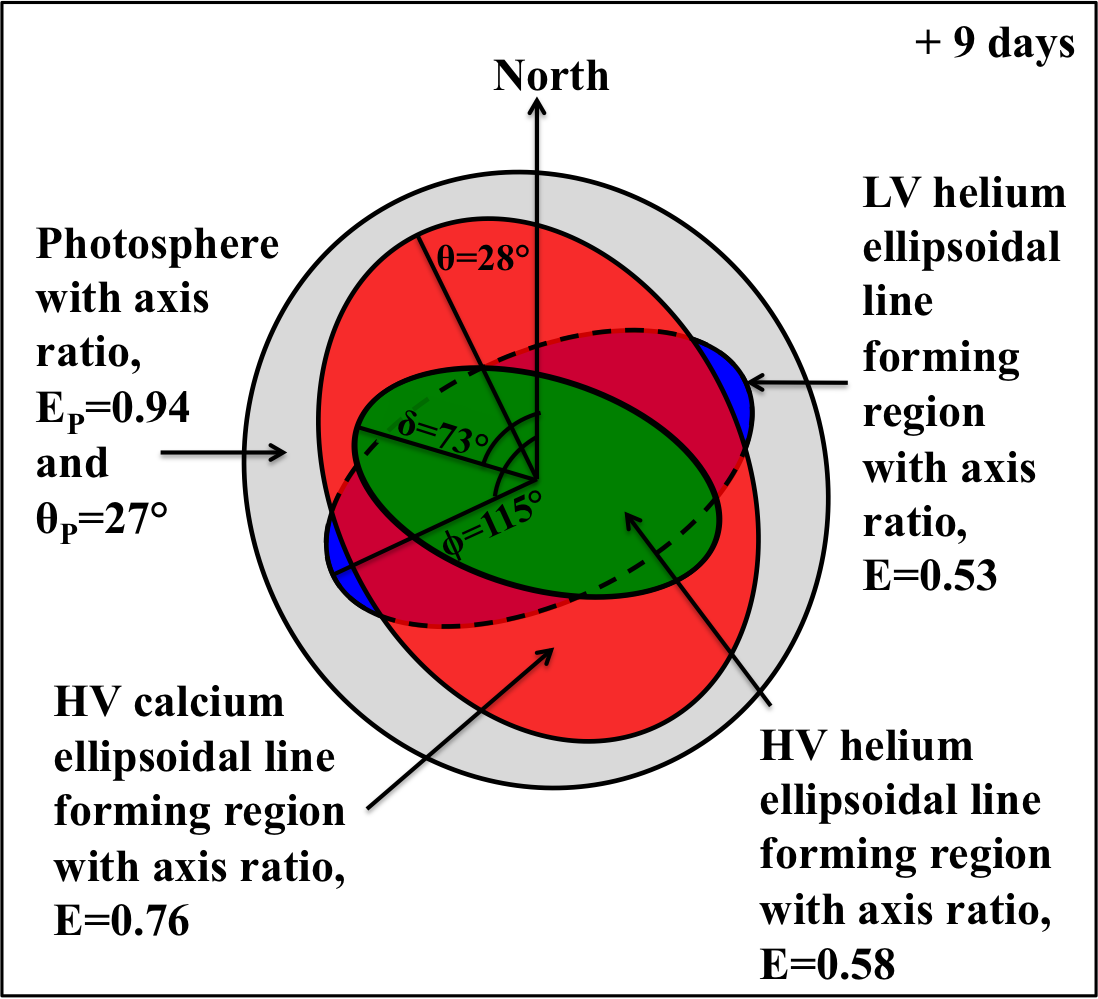}\\
\includegraphics[scale=0.425]{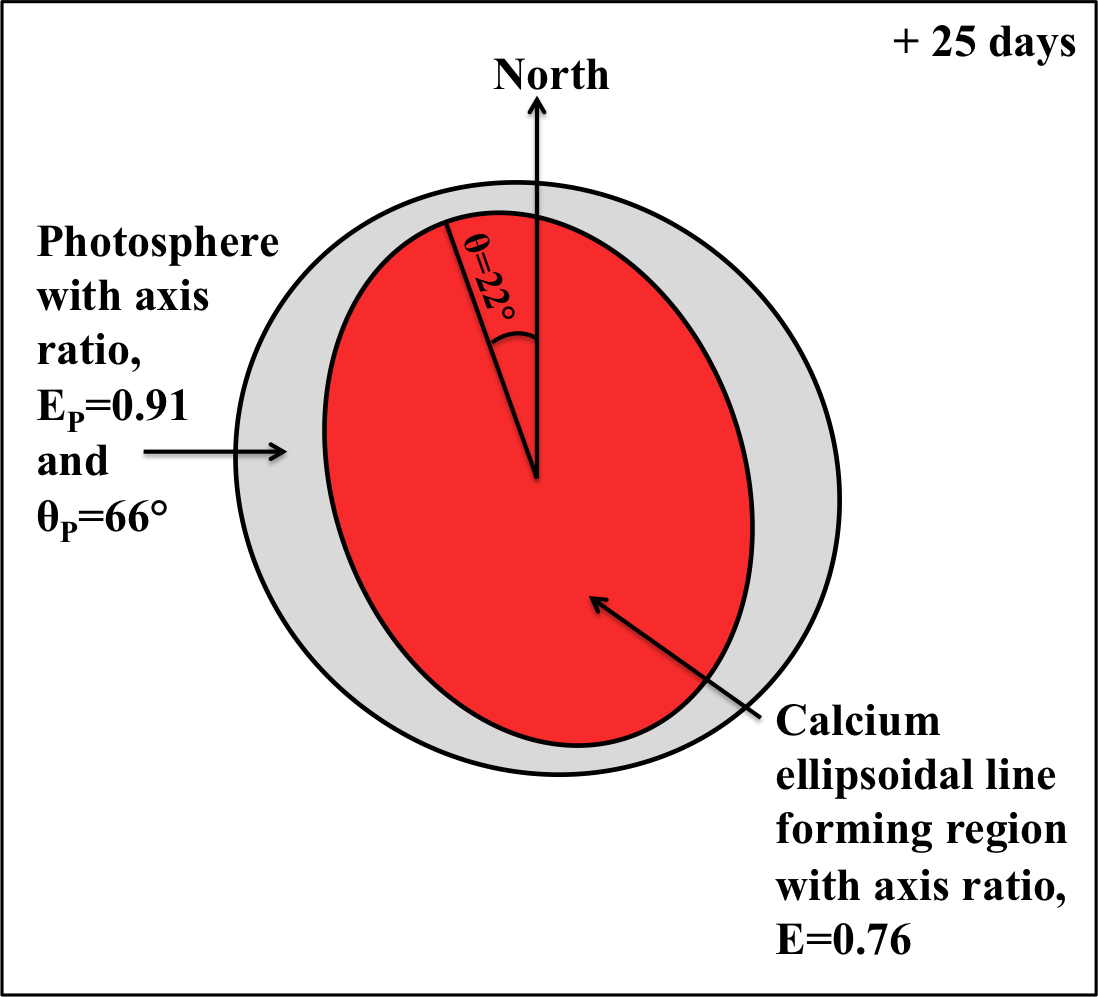}\\
\caption[]{A schematic of the simulated line forming regions at $-$10 (top), $+$9 (middle) and $+$25 days (bottom) for Ca\,{\sc ii} (red) and LV He\,{\sc i} (blue) and HV He\,{\sc i} (green) covering an ellipsoidal photosphere with an axis ratio, E$_{\rm P}$ and aligned with the long axis at $\theta_{\rm P}$ from North (grey). The polarisation simulation was able to successfully reproduce the observed continuum and line Stokes vectors.}
\label{fig:simpic}
\end{figure}

In an effort to constrain the geometry we constructed a toy model following the procedure of \citet{Mau2010}.  This model was designed to compare the observed polarisation of the blue-shifted absorption components of P Cygni profiles with the polarisation induced by blocking of the continuum light by simple obscuring line-forming regions. The Monte Carlo simulation used $1\times10^{7}$ photons, distributed across a 2D elliptical photosphere with axis ratio E$_{\rm P}$. The simulation included a spherical limb darkening model, mapped to the ellipse. Each photon was polarised according to a probability function that assigned a polarisation angle based on the position of the photon's origin from the centre of the photosphere. The photon was either assigned a random polarisation angle or a polarisation angle tangential to the ellipse at the position of the photon. Those photons closer to the limb of the photosphere were less likely to be randomly polarised. This configuration replicated the effect that photons arising from the limbs are more likely to have been scattered by 90 degrees into the line-of-sight and, hence, are more polarised. The polarisation probability function was scaled to reproduce the degree continuum polarisations predicted for pure electron scattering, oblate ellipsoids by \citet{Hof1991}.  We determined that 16.5\% of limb photons were required to be polarised parallel to the photosphere edge to reproduce the expected levels of polarisation.

Simple line-forming regions were then placed over the photosphere and assumed to have large optical depth, so as to completely absorb all photons originating from the area of the photosphere covered by the line forming region. The shapes of the blocking regions used include a circle, an ellipse with varying axis ratio, rectangular boxes (such as an edge-on disk-like CSM would appear) and triangular lobes (such as bipolar outflows might appear, as viewed from the equator). The blocking regions were rotated and/or moved across the photosphere to model the degree of polarisation, the polarisation angle and the depth of the absorption component in the flux spectrum. Combinations of shapes were also modelled simultaneously to recreate more complex line forming regions.  The final total Stokes parameters for each model were calculated following the same procedures used to determine the Stokes parameters from observational data. The fraction of photons absorbed by the line forming region was calculated and compared to that absorbed in one 15 \ang\,\,bin in the observed flux spectrum.   

This procedure was used to constrain the shape of the photosphere and principal line forming regions at 3 characteristic epochs: $-$10, $+$9 and $+$25 days.  These epochs were chosen as they probe the outer and inner most ejecta, and the transition between the two regimes.
At $-$10 and $+$25 days, we only modelled the polarisation of \canir\,.  At $+9$ days, the data show the clearest signals of separate HV and LV components of \canir\,and He\,{\sc i}, implying that any blending between the two components in the line is minimal and so the observed polarisation properties and line depth at a given velocity is solely due to one of the two components.  The model was used to reproduce the observed line depth and polarisation properties at the single wavelength at which the maximum degree of polarisation for a given line of interest was observed in our data.  Possible line forming region configurations that could reproduce the observed Stokes parameters were found by trial and error (see Figure \ref{fig:simpic}). For all lines being considered at each of the three epochs, we found that the line polarisations could be replicated by assuming a simple ellipse for each region centred on the centre of the photosphere. These ellipses were characterised by the size of the major axis of the ellipse (relative to the major axis of the photosphere), the major to minor axis ratio and the orientation angle of the major axis of the ellipse, measured East of North.

For the data at $-10$ days, we found that a configuration of an ellipsoidal photosphere with E$_{\rm P}$=0.91, with the major axis oriented at $\theta_{\rm P}=$119$^{\circ}$ reproduced the degree and angle of the continuum polarisation.  Simulations of the data
at  $+$9 and $+$25 days required the photosphere to be rotated to  $\theta_{\rm P}=$27$^{\circ}$, but with a similar axial ratio as used for the earlier epoch.  

At $-$10 days, the polarisation of \canir\, was best reproduced with an ellipse with an axis ratio of 0.71 and major axis oriented at  $\theta\sim$160$^{\circ}$, offset by 41$^{\circ}$ from that of the photosphere. In terms of axis ratio, the simulated HV Ca\,{\sc ii} did not undergo a significant change between $-$10 and $+$9 days,  however the orientation of the region rotated by $\sim$130$^{\circ}$ (or $-$50$^{\circ}$) to $\theta\sim$30$^{\circ}$.  In order to recreate the larger fraction of photons absorbed by this line at $+9$ days, the size of the obscuring region was increased from 70\% to 88\% of the major axis of the photosphere.

An ellipse with a more extreme axis ratio of 0.53 was required to reproduce the observed LV component of \hena,  oriented at $115^{\circ}$ and with size 75\% of the photosphere.  This  would imply that the LV He\,{\sc i} component covers a substantially smaller portion of the photosphere than HV Ca\,{\sc ii} and that the degree of asymmetry is more extreme.   The polarisation properties of the HV component of \hena\, could also be reproduced with an elliptical line forming region with an axis ratio of 0.58. The lower absorption depth of the HV component implies that it covered less of the photosphere than the LV component, with a major axis 60\% of the size of the photosphere.  The principal axis of the HV component line forming  region is also significantly rotated from that of the HV Ca\,{\sc ii} line forming region and the photosphere by $73^{\circ}$ (or $42^{\circ}$ from the LV \hena\, component).

The simulation indicates that neither the photosphere nor the Ca\,{\sc ii} line forming regions are subject to a dramatic change in geometry between $+$9 and $+$25 days. As there is a small scatter in the polarisation angle between all species at these epochs, it is reasonable to expect that they may all have a geometry similar to that shown in Figure \ref{fig:simpic}. 

We caution that while our assumption of line forming regions composed of simple shapes can reproduce the observed levels of polarisation, it does not imply that these are unique solutions nor that they are necessarily correct.  In addition, the geometries for the line forming regions we have identified using our toy model are also dependent on the choice of the ISP. 

%%%%%%%%%%%%%%%%%%%%%%%%%%%%%%%
%%%%%%%%%%     Discussion   %%%%%%%%%%%%%
%%%%%%%%%%%%%%%%%%%%%%%%%%%%%%%

\section{Discussion}
\label{sec:discussion}
\subsection{iPTF 13bvn in the Context of Other Stripped CCSNe}
iPTF 13bvn stands out amongst the limited sample of stripped-envelope CCSNe for which there are polarimetric observations, by having such a dense time series of spectropolarimetric observations (6 epochs). The observed polarisation properties of this class of SNe are very diverse, particularly regarding the line polarisation. iPTF 13bvn shows similarities to the other stripped-core SNe, however there are also some aspects in which it exhibits significantly different behaviour.

The low levels of continuum polarisation observed here are consistent with those found in other stripped CCSNe, such as SNe 2008D, 2007gr, 2005bf, and 2002ap \citep{Mau2009, Tan2008, Mau2007b, Kaw2002}. These levels of polarisation limit the degree of global asymmetries to $\lesssim$10\% \citep{Hof1991}.  A feature consistently observed in all stripped CCSNe \citep{Mau2009}, and some thermonuclear Type Ia supernovae \citep{Maz2005}, is the absorption trough of \canir\,, which is commonly associated with high velocities ($\sim$10-20,000 \kms) and large degrees of polarisation ($\sim$2-4\%). This feature tends to form at higher velocities in the ejecta and with a polarisation angle that is markedly different to those of  other species \citep{Tan2012, Mau2009, Tan2008, Mau2007a, Mau2007b}. In addition, occasionally separate LV and HV components can be resolved in the Ca\,{\sc ii} line profile in the flux spectrum \citep[e.g.][]{Mau2009}. Species such as He\,{\sc i}, Fe\,{\sc ii}, O\,{\sc i}, Si\,{\sc ii} and Na\,{\sc i}\,{\sc d} are frequently observed to form closer to the photosphere and with similar polarisation angles that are distinct from that of Ca\,{\sc ii} \citep[e.g. SN 2008D][and their Figure 11]{Mau2009}. For iPTF 13bvn, the velocity at the \canir\,\,absorption minimum is substantially higher than that of the other species. At maximum light and a week later, the polarisation angle at maximum polarisation for \canir\, is different from He\,{\sc i} and Fe\,{\sc ii}. The multi-epoch observations here allow us to follow the temporal evolution of the polarisation associated with spectral lines. In iPTF 13bvn both Ca\,{\sc ii} and He\,{\sc i} are observed to have photospheric and HV components that are geometrically distinct. Fe\,{\sc ii} $\lambda4924$ is also polarised at similar velocities and has a similar orientation the HV He\,{\sc i} and Ca\,{\sc ii} line forming regions, which is unique to iPTF 13bvn.

The only Type Ib SNe with spectropolarimetric  observations, other than iPTF 13bvn, are SNe 2005bf, 2008D and SN 2009jf \citep{Mau2007b, Mau2009, Tan2012}. \citet{Mau2007b} presented polarimetric observations of the unusual SN 2005bf that was originally classified as a Type Ic SN \citep{Mor2005}, before the later appearance of He lines \citep{Fol2006}. For SN 2008D \citet{Mau2009} presented two epochs of spectropolarimetry at {\it $\sim $V}-band maximum and 15 days later. At the closest comparable epochs of 0 and $+$25 days, SN 2008D shows similarities to iPTF 13bvn in the polarisation associated with certain spectral lines. For SN 2008D, the velocities and polarisation angles at the absorption minimum of He\,{\sc i} and Fe\,{\sc ii} lines indicates that the two species are found in similar parts of the ejecta, while Ca\,{\sc ii} is found at higher velocities and separated in polarisation angle by $\sim$100$^{\circ}$. iPTF 13bvn shows similar behaviour to SN 2008D at 0 days, where He\,{\sc i} and Fe\,{\sc ii} are observed to have similar polarisation angles at similar velocities, while \canir\, is observed at a higher velocity with a polarisation angle offset from the other species by $\sim$70$^{\circ}$. \citet{Mau2009} also observe significant polarisation associated with the O\,{\sc i} $\lambda 7774$ line, the polarisation angle of which aligns with that of high velocity Ca\,{\sc ii} at {\it V}-band maximum. For iPTF 13bvn, however, O\,{\sc i} $\lambda 7774$ appears much later and there is no significant polarisation associated with it until $+$36 days, at which stage the polarisation angle is aligned with the continuum and all other polarised species. \citet{Mau2009} also found a polarisation maximum around 6200 \ang\,\,which they attribute to a blend of Si\,{\sc ii} and a high velocity component of \halpha\,\,at $-$17,050 \kms.  Similarly to iPTF 13bvn, the feature had also disappeared by the later epoch.

\subsection{Complex Multi-axis Symmetry}

The spectropolarimetric data of iPTF 13bvn have revealed the presence of multiple separately polarised components to the ejecta. The observations of loops on the $q$-$u$ plane indicate that the orientation of the line forming regions for \canir and \hena\, change with velocity (radius).  The loops observed at $+$9 days and the step-like discontinuities in the polarisation angle across the \canir\, and \hena\ profiles at $+$7 and $+$9 days imply that there are two geometrically distinct line forming regions for the LV and the HV components.

The LV components of He\,{\sc i} and \canir\, have the same polarisation angle, implying that they form in the same location in the ejecta.  These separately polarised components may suggest either two physically distinct line forming regions for both calcium and helium or a change in the underlying ionisation structure \citep{Chu1992}. The possible geometry of LV He\,{\sc i}, from the Monte Carlo simulations, may then also be applicable to LV Ca\,{\sc ii}.  Although we have assumed simple ellipsoidal configurations for the line forming region, and have not explored more complex geometries, we can rule out some geometric configurations.  The rotation of the polarisation angle  across these lines, compared to the continuum polarisation angle, shows that a shell with a similar shape to the photosphere that selectively blocks only part of the unpolarised light cannot be responsible for the observed degree of polarisation \citep{Kas2003}.   The small scatter in the polarisation angle across the LV components also suggests that this shared line forming region is a single continuous structure in the ejecta, rather than arising in clumps \citep{Mau2010, Hol2010}. In terms of the simple geometries explored in the simulation here and those presented by \citet{Kas2003} and \citet{Hol2010}, a non-clumpy elliptical line-forming region or shell, with different orientations to the asymmetry of the photosphere, can provide a reasonable explanation for the observed polarisation properties of iPTF 13bvn. Previously, \citet{Mau2007b} inferred spherical configurations for the LV line-forming regions for SN 2005bf. In addition, although the jet-torus paradigm \citep{Kho1999} has been invoked to explain spectropolarimetric observations of some stripped CCSNe \citep{Tan2008},  we can exclude a torus-like line-forming region for the LV components as we do not observe the high levels of polarisation predicted for such a geometry \citep{Kas2003}.
 
The HV component of \canir\, in iPTF 13bvn  is observed to be highly polarised, with a velocity much greater than the bulk of the ejecta. For iPTF 13bvn, we observe the same polarisation angle in HV Ca\,{\sc ii}, HV He\,{\sc i} and Fe\,{\sc ii}, which is in stark contrast to the distinctly different polarisation angle for \canir\,, compared to other species, observed for SNe 2005bf, 2007gr and 2008D \citep{Mau2007b, Tan2008, Mau2009}.  Similarly to the LV component, the polarisation properties of HV Ca\,{\sc ii} were best replicated in the polarisation simulation with an elliptical line-forming region.  It is interesting to note that the presence of two velocity components for \canir\, is only discernible in the spectropolarimetric data, and that these two components are not resolved in the line profile of the flux spectrum.  This undermines the accuracy of our simulation, which assumes that the line depth and polarisation were due to the same line forming region.  The 2D bipolar model of \citet{Tan2012} produces a straight line on the $q$-$u$ plane which also provides a match to the observed polarisation characteristics of HV Ca.  It is unclear to what degree \naid\, plays a role in the \hena\, blend, and therefore whether He\,{\sc i} and Ca\,{\sc ii} share the same line-forming region. The results of the polarisation simulation suggest that, likewise, an elliptical line forming region can also reproduce HV He\,{\sc i} but that is rotated compared to HV Ca\,{\sc ii}.

In the later epochs, ($+25$ and $+36$ days) the loops on the $q$-$u$ plane across the \hena\,\,blend and  \canir\,\,profile (Figure \ref{fig:quline}) appear more variable with several peaks in the degree of polarisation apparent in Figure \ref{fig:wlpol}. The evolution across the line is reminiscent of those generated in the simulations by \citet{Hol2010}, where the ejecta are composed of clumps separated in velocity space. This suggests that the ejecta, while apparently still following a principal axis of symmetry, develop a degree clumpiness at later times. At $-10$ days, the line polarisation of the LV and HV component share similar polarisation angles consistent with a single axial symmetry, with no evidence of clumpiness within the ejecta. The implication is that between $+$9 and $+$25 days the ejecta became clumpy, most likely due to hydrodynamic instabilities. Such instabilities could propagate through the ejecta \citep{Blo2001} or may develop from remnant asphericities in the explosion mechanism itself, such as the  standing accretion shock instability \citep{Jan2007} or through the propagation of a jet-like flow through the progenitor star \citep[][see also \citealt{Whe2008} for discussion of jet-induced hydrodynamical instabilities in SNR Cas A and \citealt{Ham2010} for simulations of instabilities in supernova explosions]{Cou2009}.

There are no previous observations of two geometrically and physically distinct He\,{\sc i} line forming regions in the ejecta of Type Ib SNe. The properties of both the HV and LV components for iPTF 13bvn are different from those observed for other stripped CCSNe. It is possible that the scarcity of previous polarimetric observations or viewing angle effects may have meant that this feature has been missed in the other SNe. If these features are truly unique to iPTF 13bvn, signs of the SN's peculiarity may be expected in the light curve or spectra. iPTF 13bvn was classified as a Type Ib SNe, with similarities to SN2009jf, while the light curve was observed to be fast declining and of low luminosity \citep{Sri2014}. It is possible that the observations of the two He\,{\sc i} components are related to the plateau in the evolution of the He\,{\sc i}\, velocity (see Fig. \ref{fig:minabsvel}). A similar plateau is also observed in SN 2008D \citep[see][and their Figure 13]{Mod2009}, however the polarised HV and LV components in Ca\,{\sc ii} and He\,{\sc i} observed in iPTF 13bvn are not seen in SN 2008D \citep{Mau2009}. Only increasing the sample of Type Ib SNe with high polarimetric coverage can resolve the question of whether this complex ejecta morphology is common to all Type Ib SNe or only a subset, and what role the explosion mechanism plays in the formation of these multiple components.

\subsection{Non-thermal excitation}
The velocity at the absorption minimum of the \hena\, blend is observed to decay steeply before plateauing at around maximum light (see Figure \ref{fig:minabsvel}).  The velocities measured for the He\,{\sc i} $\lambda6678,7065$ lines also appear to plateau at the same time \citep[see][and their Figure 6]{Fre2014}. The plateau in the He\,{\sc i} velocity may arise from non-thermal excitation, from the direct deposition of gamma rays from the radioactive decay of $^{56}$Ni and $^{56}$Co, at velocities higher than the photospheric velocity \citep{1991ApJ...383..308L}. The geometry of He\,{\sc i} may, therefore, serve to trace the geometry of the radioactive nickel.  The coincident emergence of the HV and LV components in the polarimetry with the beginning of the He velocity plateau at  $\sim$ maximum light suggests that the non-thermal excitation of the He\,{\sc i} lines also powers those components. 

In principle, the collision of an approximately spherical ejecta with an aspherical circumstellar medium (CSM) could also produce the observed early evolution in the polarimetry. If interaction were the dominant source of non-thermal excitation, we might expect to also see narrow emission lines in the spectrum, which are not seen in our observations of iPTF 13bvn.  Furthermore, the collision of ejecta and a CSM would result in an injection of energy, identifiable in the light curve as a decline that is slower than expected from radioactive decay. \citet{Sri2014} note, however, that iPTF 13bvn has a faster decline than most Type Ib/c SNe and faster than expected for the decay of $^{56}$Co. This suggests a deficiency of energy in the ejecta due to incomplete trapping of gamma rays as opposed to energy from interaction with the CSM.   The observations, therefore, disfavour interaction as the source of the separate components and the He\,{\sc i} velocity plateau and are, instead, consistent with $^{56}$Ni located above the photosphere at and after the light curve maximum.  The photospheric velocity measured at $18$ days post-explosion reveals that nickel must be mixed out to $\sim$50-65\% of the ejecta, by radius, at $\sim$8700 \kms. \citet{Ber2014} performed one dimensional hydrodynamical explosion models of iPTF 13bvn which required $^{56}$Ni to be mixed out to $\sim$96\% of the initial mass of the progenitor to reproduce the observed rise time of the light curve. The discrepancy between these two estimates for the degree of Ni mixing most likely arise in the application of a 1D model assuming spherical symmetry to, as we have observed, an asymmetrical situation. If the polarisation of the HV and LV He\,{\sc i} components at $+$7 and $+$9 days trace the 2D projected distribution of nickel, this implies that not only is it not spherically distributed, covering only a portion of the photosphere, but also that it is distributed in at least two (if not more) outflows originating from the core.  Our observations have demonstrated the importance of spectropolarimetry for measuring how Ni is mixed into the outer layers of the SN ejecta and the degree of this mixing.  This has important implications for how photometric and spectroscopic observations are interpreted, especially in the context of 1D models.

\subsection{Implications for the Explosion Geometry}
 The absence of any strong O\,{\sc i} lines in the spectrum until $+$25 days indicates that the core is still shielded by the photosphere until that epoch.  The presence and polarisation of He {\sc i}, at $\sim$ maximum light, indicates the role of non-thermal excitation from Ni deposited above the core by an axisymmetric outflow that stalled in the helium mantle. 

The approximate alignment of the continuum polarisation angle in the later epochs ($+25$ and $+36$ days) with that of the HV component suggests that there is a principal axis of symmetry for the explosion which may lend itself to the interpretation of a bipolar or jet driven explosion. In this model a jet, originating either from asymmetric neutrino emission \citep{Blo2007} or the magneto-rotational mechanism \citep{LeBlancWilson1970, Mik2008}, would carry heavy elements in bipolar flows, while low and intermediate mass elements (e.g. helium and oxygen) would form a torus in the equatorial plane \citep{Kho1999,KhoHof2001, Mae2002}. We observed no such segregation between the locations of heavier mass (iron and calcium) and lower mass (helium and oxygen) elements in the ejecta, that would suggest such a geometric configuration for iPTF 13bvn.  The presence of a separate LV component of He\,{\sc i}\,, which must also arise from non-thermal excitation, also implies that the distribution of nickel cannot be described by just a simple bipolar geometry.

In comparison with the velocity of the HV He component ($\sim-$8700 \kms) which we use as a tracer for the Ni distribution, the high velocities ($\sim-$13,000 \kms) measured for Fe\,{\sc ii} lines at early times implies these features arise from primordial iron already present in the envelope of the progenitor. Likewise, the low velocity inferred for the Ni makes it improbable that the \canir\, HV component traces a fast moving jet.  The deposition of nickel outside of the photosphere, at $\sim$ maximum light, would lead to significant losses of gamma ray photons, which would otherwise be trapped in the ejecta under the photosphere.  This could explain the low luminosity and fast light curve decline of iPTF 13bvn \citep{Sri2014}, and may also imply that the nickel mass derived from the bolometric light curve \citep[0.05-0.1\msol][]{Sri2014, Ber2014} may underestimate the true value. 

Previously, the action of bipolar flows has been inferred from spectropolarimetry of other CCSNe. \citet{Tan2008} observed HV Ca in the Type Ic SN 2007gr, which they proposed was formed in a bipolar outflow, with oxygen distributed in a torus in the equatorial plane. 
Polarisation in the wavelength region around the \canir\, in the Type Ic SN 2002ap was interpreted by \citet{Kaw2002} as indicating the presence of a relativistic jet (0.115$c$) that had punched through the ejecta.  Later \citet{Wan2003ap} presented 3 epochs of spectropolarimetry of SN 2002ap and challenged this interpretation, suggesting the polarisation was associated with oxygen at much lower velocities and suggested the data were consistent with a bipolar jet-like flow that had stalled in the core.  Similarly, \citet{Mau2009} invoked a thermal energy dominated jet-like flow, that had stopped within the core, to explain the orthogonality between the polarisation angle of HV Ca\,{\sc ii} with the angles of helium and iron at lower velocities in the Type Ib SN 2008D. \citet{Mau2007b} also favoured a jet, that stopped above the core in the helium mantle, for the Type Ib/c SN 2005bf. \citet{Mau2015} also favour plumes of radioactive nickel to explain the polarisation of the Type IIb SN 2011dh. \citet{Mau2009} and \citet{Wan2003ap} predicted that, as the photosphere recedes through the deeper layers, later observations of Type Ibc SNe would reveal evidence of an asymmetric flow originating from the explosion. Our multi-epoch spectropolarimetry of iPTF 13bvn has confirmed these predictions.  Although there are differences between iPTF 13bvn and these other Type Ibc SNe, overall a consistent picture of the geometry of these events is emerging.

The geometrical properties inferred here are reminiscent of the complex morphology of the Galactic Type IIb SN remnant Cas A. The presence of the opposing North East ``jet'' and South West ``counter-jet'' are indicative of axisymmetric bipolar flows associated with the explosion, similar to that inferred here for the observations at later epochs. Several authors agree, however, that these so-called ``jets'' are secondary features, perhaps arising due to instabilities \citep{Whe2008}, accretion onto a neutron star \citep{Jan2005} or due to precession of the axis of rotation when the bipolar flows were launched \citep{Bur2005}. Similarly, \citet{Mil2013} conclude that while the observed jets are unlikely to be the result of a jet-induced explosion they are intrinsic to the explosion. \citet{Whe2008} proposed a model in which the jet-induced explosion launched products from explosive nucleosynthesis in a jet in the South East direction, not along the NE/SW axis.  \citet{Blo2001bub} and \citet{Mil2013} interpreted the presence of several large ejecta rings surrounding iron rich ejecta to be due to the non-axisymmetric deposition of nickel in clumps in the explosion and such a scenario has also been proposed for SN 1987A \citep{Li1993}. The inference of two outflows of nickel and additional ``clumps'' or ``bubbles" of nickel in iPTF 13bvn may therefore have very real analogues in resolved SN remnants like Cas A \citep{2015Sci...347..526M}.

%%%%%%%%%%%%%%%%%%%%%%%%%%%%%%%%%%%%%%%%%%%%%%%%%%%%%%
%%%%%%%%%%%%%%%%%%%%%%%%%%%%%%%%%%%%%%%%%%%%%%%%%%%%%%
\subsection{Polarimetric Evolution and the Implications for the Progenitor System}
%%%%%%%%%%%%%%%%%%%%%%%%%%%%%%%%%%%%%%%%%%%%%%%%%%%%%%
%%%%%%%%%%%%%%%%%%%%%%%%%%%%%%%%%%%%%%%%%%%%%%%%%%%%%%

The possible identification of a progenitor candidate for iPTF 13bvn, as reviewed in Section \ref{sec:intro}, has generated debate over the progenitor system that gave rise to this SN.   For iPTF 13bvn, we have identified 3 main axes of symmetry over the period covered by our spectropolarimetric observations corresponding to: the HV component at $+$9 days and the line and continuum polarisation in the later epochs at $\sim$135$^{\circ}$; the LV component at +9 days at $\sim$25$^{\circ}$ degrees; and the alignment of the line polarisation angles of multiple species at the earliest epoch of $-$10 days at $\sim$75$^{\circ}$. The asymmetries in the earlier epochs reflect the shape of the the outer most layers of the ejecta. The difference in the polarisation angles observed at $-$10 days and those at later epochs ($+$25 and $+$36 days) suggests that the outer layers of the ejecta did not share the same geometry as the explosion. This suggests, similarly to what has been observed for Type IIP SNe arising from red supergiant progenitors \citep{Leo2006, Chor2010}, that the envelope of the pre-supernova star must have been sufficiently large in order to conceal the inner explosion asymmetries for 18 days (until 0 days when evidence of the inner explosion geometry began to emerge).

In the context of understanding the origin of the geometry of the outermost ejecta, the identification of the polarised absorption feature at $\sim$6200 \ang\,\,becomes especially interesting. The feature is significantly polarised in the first two epochs only ($-$10 and 0 days) before both the degree of polarisation and absorption depth decrease. As has been suggested for other Type Ib SNe \citep{Whe1994, Den2000, Bra2002, Par2015}, if this feature is H$\alpha$ at $-$16,400 \kms, this would imply that the H line-forming region lies in the outermost layers of the ejecta, perhaps in a shell at the outer edge of the ejecta, as was suggested for Type Ib SN 2008D \citep{Mau2009}. The polarisation of this feature may then reflect asymmetries in the outermost layers of the progenitor or the progenitor's wind swept up by the ejecta, as has been proposed for some Type IIb SNe. The polarisation angle of the 6200 \ang\,\,feature is consistent with those of the rest of the species at $-$10 days, in contrast to the \halpha\,\,polarisation of Type IIb SN 2001ig, where it is found to have a different distribution within the ejecta compared to the dominant axis \citep{Mau2007}. Similarly, \citet{Cho2011} found that the \halpha\,\,polarisation was aligned with that of \hefive\,\,but rotated with respect to Ca\,{\sc ii} and Fe\,{\sc ii} for transitional Type IIb/Ib SN 2008ax. In contrast to the modest polarisation of $\sim$0.6\% found here, \citeauthor{Cho2011} determined large variations of $\sim$3\% across \halpha, indicating large asphericities in the outer ejecta. SNe 1993J and 1996cb showed more modest polarisation across \halpha\,\,of $\sim$1.5\% \citep{Tran1997, Wan2001}. For SN 2011dh, \citet{Mau2015} found that the polarisation of enhanced line polarisation associated with \halpha\,\,peaked at 30 days post-explosion and aligned with the continuum polarisation at early times. This, however, was not related to the shape of the outer layers of the ejecta but due to plumes of $^{56}$Ni deposited outside of the photosphere according to \citeauthor{Mau2015}. In general, the polarisation properties of \halpha\,\,in Type IIb and Ib SNe are quite different which may imply distinctly different progenitor configurations for the two types of SNe.

The shape of the outer layers of the SN ejecta is also subject to the geometry of the shock break out, which is dependent upon the geometry of the explosion and dynamics of the interaction of the shock front as it propagates through the progenitor star. The post-maximum polarisation properties indicate that the explosion of iPTF 13bvn was highly asymmetric. \citet{Mat2013} performed calculations which showed that the resulting shock front of an aspherical explosion may emerge at an oblique angle to the stellar surface. A key prediction of these models is this would induce large asphericities in the highest velocity material and hence significant polarisation. At later times the polarisation angles are then predicted to rotate, as the geometry becomes progressively dominated by the shape of the explosion mechanism, as we have observed for iPTF 13bvn.  The emergence of the explosion geometry at $0$ days (18 days post-explosion) implies the progenitor of iPTF 13bvn must have had a relatively extended envelope, which argues against oblique shock break out as the source of the early polarisation that \citet{Mat2013} determined is more easily produced by compact progenitors, such as WR stars.

The early polarisation may also arise in a distorted stellar envelope. Tidal distortion of the progenitor star as a result of binary interaction has been discussed as the source of the observed polarisation in numerous Type IIb SNe  including SN 1993J and recently in SN 2011dh by \citet{Mau2015}. \citet{Tran1997} and \citet{Tra1993} observed high levels of continuum polarisation in SN 1993J, of $p\sim$1-2\% which exhibited no strong temporal dependence. \citet{Hof1995} found that this could be produced by an oblate model with an axis ratio of 0.6, while \citeauthor{Tran1997} and \citeauthor{Tra1993}  argued that the lack of strong temporal change in the polarisation was an indication that the asymmetry was not intrinsic to the explosion but due to a distorted envelope. The strong asphericity was linked to a pre-SN common envelope phase in a close binary system \citep{Hof1995}. Similar polarisation properties were observed for SN 1996cb \citep{Wan2001} and SN 2001ig \citep{Mau2007}, with \citeauthor{Mau2007} suggesting that they were the result of a similar binary progenitor system. The alignment of the \canir, Fe\,{\sc ii} and Si\,{\sc ii} polarisation angles and the non-zero degree of continuum polarisation ($\sim$0.2-0.7\% depending on the chosen ISP) suggests that the envelope was distorted in an axisymmetric manner. The effect of the tidal distortion of the progenitor on the early-time polarisation of the SN may be amplified by how the explosive shock is able to propagate through the distorted star.  Fast rotation or tidal interaction with a binary companion leads to a density profile in the progenitor that is steeper along the direction of the poles. \citet{Ste1992} showed that a spherical shock front would take on a prolate shape upon encountering the steeper density gradient along the poles.  For a massive, differentially rotating star (like the progenitor of SN 1987A), \citeauthor{Ste1992} found that at $\sim 30$ days post explosion the photosphere receded through the oblate ejecta to the prolate regions, producing a rotation of the polarisation angles. The observations of iPTF 13bvn are also roughly consistent with this model.

Spectropolarimetric surveys of galactic and Large Magellenic Cloud (LMC) Wolf-Rayet stars show that the majority ($\sim$80\%) are slow rotating spherically symmetric stars \citep{Har1998, Vin2007,StL2011}. Based on geometry alone, the progenitor of iPTF 13bvn is unlikely to have been a slowly rotating WR star. \citet{Vin2011} and \citet{Gra2012} discovered that a subset of WR stars that were rotating sufficiently fast to result in a flattened wind were also associated with ejecta nebulae, such that they concluded that these had to be young WR stars (having recently been through a luminous blue variable or red supergiant phase).  In low metallicity environments the line-driven winds of these fast rotators will be too weak to remove significant amounts of angular momentum \citep{YooLan2005}, such that they would retain any rotational induced asymmetries up to the point of explosion. \citet{Vin2007} proposed that this effect was at most limited to metallicities below that of the LMC.  \citet{Kun2015} measured the metallicity of the region of iPTF 13bvn as being slightly subsolar, which would suggest as WR star in this region would not be expected to explode as a distorted, fast rotator.  

The polarimetric observations may be more easily explained in the context of the close binary model proposed by \citet{Ber2014}. The explosion of a low mass He star in a binary system does not require the compactness implied by the explosion of a single massive WR star. Indeed \citet{Ber2014} showed that the non-detection of shock cooling in the early light curve \citep{Cao2013}, does not preclude the existence of an extended envelope and find the radius of the progenitor may be as large as R$\lesssim$150 \rsol. \citet{Kim2015} also find that the pre-explosion photometry can be replicated with a helium star progenitor of 3.0-4.4 \msol\,\,(with an O-type star companion) and extended radius of $\sim$12-30 \rsol. Interestingly, stars in this mass range are expected to undergo mass transfer (Case BB) in the final decades before core collapse \citep{Yoo2010}. In this case, the extended helium envelope of the star would be significantly distorted, having filled a Roche Lobe that deviates dramatically from spherical symmetry. In addition, stars in this mass range (3-3.8 \msol) are also predicted to retain a residual amount of hydrogen in the outermost layers \citep{Yoo2010}. Synthetic spectra of SNe resulting from the explosion of stars in this range exhibit H\,{\sc i} lines at early times up until peak \citep{Des2015}, similarly to the absorption feature at $\sim$6200 \ang\,\,observed in iPTF13bvn. The polarimetry may then support the explosion of a star with final mass of 3-3.8 \msol\,\,with a recent binary interaction phase. Questions remain, however, as to whether the distortion of the helium envelope from mass transfer in the previous decades would be preserved up until core collapse and through to the observations at $-$10 days. Nevertheless, the mass transfer would result in a complex CSM from which the early polarisation may result. It should be noted that mass transfer during carbon-oxygen core contraction with residual hydrogen remaining in the envelope is predicted only within a narrow mass range. Should the progenitor mass be several times higher than the estimates of a few solar masses, which \citet{Whe2015} suggest could be the case, the evolutionary paths will be vastly different and a significantly distorted Roche Lobe in the final decades of the star's life may not be expected.

In summary, our polarimetric observations favour an extended, distorted progenitor (possibly with residual H in the outer layers) rather than a compact WR star.

\section{Conclusions}
We find iPTF 13bvn to have exhibited significant continuum and line polarisation in observations covering the period from $-$10 to $+$36 days, with respect to the {\it r}-band light curve maximum. The continuum was intrinsically polarised at the 0.2-0.4\% level throughout the observations, implying asphericities of $\sim$10\% in the shape of the photosphere. The degree and angle of continuum polarisation was found to be dependent on the choice of the ISP.

Significant polarisation was found to be associated with the absorption profiles of spectral lines of \canir, \hena, He\,{\sc i}\,$\lambda\lambda 6678, 7065$, Fe\,{\sc ii}\,$\lambda 4924$, Si\,{\sc ii}\,$\lambda 6355$ and O\,{\sc i}\,$\lambda 7774$. The \canir\,\,absorption showed the strongest degree of polarisation of $\sim$3\% at $+$25 days.  The polarisations of the \hena\,\,and Fe\,{\sc ii} reached maximum levels of $\sim$1\%, followed by O\,{\sc i} and Si\,{\sc ii}/\halpha\,\,at $\sim$0.6\%.  We propose that the observation of two loops on the $q$-$u$ plane associated with the \hena\,\, profile at $+$7 and $+$9 days indicates the presence of at least two physically distinct components to the ejecta. The formation of separate loops due to the blending of Na\,{\sc i}\,{\sc d} with He\,{\sc i} was considered unlikely due to the sharp discontinuity observed in the polarisation angle across the profile and the lack of evidence of a strong contribution to the blend from Na\,{\sc i}\,{\sc d} at $+$7 and $+$9 days.  Both loops are attributed to He\,{\sc i}.  Similarly, Ca\,{\sc ii} also show a ``step-function''-like trend in the evolution of the polarisation angle with velocity, supporting the interpretation of at least two distinct line forming regions for He\,{\sc i} and Ca\,{\sc ii} in the ejecta. The LV components of Ca\,{\sc ii} and He\,{\sc i} were found to share the same polarisation angle implying a shared line forming region for the two ions.  The HV components were observed with polarisation angles separated by $\sim$40$^{\circ}$. 

The simultaneous emergence of the HV and LV components in the polarisation data with the onset of the He velocity plateau indicated the presence of non-thermal excitation. The polarisation of the HV and LV He\,{\sc i} components was used to trace the 2D projected distribution of nickel in the ejecta, and suggests the presence of at least two outflows originating from the core.  We are able to constrain the bulk of nickel to lie interior to $\sim$50-65\% of the ejecta radius, at a velocity of $\lesssim$8700 \kms. This is in disagreement with the prediction from 1D hydrodynamical models, requiring mixing  out to $\sim$96\% of the initial mass of the progenitor to reproduce the observed rise time of the light curve. 

The approximate alignment of the continuum polarisation angle in the later epochs with that of the HV component suggests that there was a main axis of symmetry in the explosion, which could be consistent with a bipolar or jet-driven explosion.  The high quality of our data reveal, however, that the structure of the ejecta is not just described by a simple, single axial symmetry and that the morphology of the explosion is much more complicated. Other outflows, originating from the explosion, appear to have been launched in multiple directions but, similarly to other Type Ib SNe, stalled in the He mantle of the progenitor star. The structure of the outer layers of the ejecta are consistent with a rotationally supported, non-WR star that had retained a thin shell of H at the time of explosion.  Our spectropolarimetric observations of iPTF 13bvn have revealed a complex explosion geometry that is reminiscent of that observed in the Galactic SN remnant Cas A.

\section{Acknowledgements}
The authors would like to thank the anonymous reviewer for their careful examination of this manuscript and the comments which helped to improve it. We are also grateful to Prof. S.-C. Yoon for helpful comments on a draft of the manuscript. ER is supported through a PhD studentship awarded by the Department of Education and Learning of Northern Ireland. The research of JRM is supported through a Royal Society University Research Fellowship. JCW is supported by NSF Grant AST 11-09881. JMS is supported by an NSF Astronomy and Astrophysics Postdoctoral Fellowship under award AST-1302771.  The authors are grateful to the ESO for the generous allocation of observing time. They especially thank the staff of the Paranal Observatory for their competent and never-tiring support of this project in service mode.

\bibliographystyle{mn2e}

\end{document}